%% file: main.tex
\documentclass[11pt,a4paper]{article}

\usepackage[T1]{fontenc}
\usepackage[utf8]{inputenc}
\usepackage{lmodern}
\usepackage{microtype}
\usepackage[margin=2.5cm]{geometry}

\usepackage{amsmath}
\usepackage{amssymb}
\usepackage{amsthm}
\usepackage{mathtools}
\usepackage{bm}

\usepackage{graphicx}
\usepackage{booktabs}
\usepackage{multirow}
\usepackage{array}
\usepackage{caption}
\usepackage{subcaption}

\usepackage{algorithm}
\usepackage{algpseudocode}

\usepackage{authblk}

\usepackage[numbers,sort&compress]{natbib}
\usepackage{xcolor}
\usepackage[
    colorlinks=true,
    linkcolor=blue,
    citecolor=blue,
    urlcolor=blue
]{hyperref}
\usepackage[nameinlink,capitalise]{cleveref}

\usepackage{enumitem}
\setlist{nosep}

\newtheorem{proposition}{Proposition}

\theoremstyle{definition}

\theoremstyle{remark}

\newcommand{\ket}[1]{\left|#1\right\rangle}

\newcommand{\norm}[1]{\left\lVert#1\right\rVert}
\newcommand{\abs}[1]{\left|#1\right|}

\newcommand{\sign}{\operatorname{sgn}}

\usepackage[mathlines]{lineno}
\newcommand*\patchAmsMathEnvironmentForLineno[1]{%
  \expandafter\let\csname old#1\expandafter\endcsname\csname #1\endcsname
  \expandafter\let\csname oldend#1\expandafter\endcsname\csname end#1\endcsname
  \renewenvironment{#1}%
    {\linenomath\csname old#1\endcsname}%
    {\csname oldend#1\endcsname\endlinenomath}}%
\newcommand*\patchBothAmsMathEnvironmentsForLineno[1]{%
  \patchAmsMathEnvironmentForLineno{#1}%
  \patchAmsMathEnvironmentForLineno{#1*}}%
\AtBeginDocument{%
  \patchBothAmsMathEnvironmentsForLineno{equation}%
  \patchBothAmsMathEnvironmentsForLineno{align}%
  \patchBothAmsMathEnvironmentsForLineno{flalign}%
  \patchBothAmsMathEnvironmentsForLineno{alignat}%
  \patchBothAmsMathEnvironmentsForLineno{gather}%
  \patchBothAmsMathEnvironmentsForLineno{multline}%
}

\title{
    Sparse Walsh-Pauli Correlation Encoding for MaxCut:
    A Quantum-Inspired Formulation with Quantum-Hardware Validation
}

\author[1,2]{C. A. do Amaral\thanks{Corresponding author: \texttt{cesar.amaral.BE@eldorado.org.br}}}
\author[3]{N. Legnazzi}
\author[1]{M. V. Reballo}
\author[1,4]{M. Ritt}
\author[1]{A. S. da Rosa Júnior}
\author[1]{F. A. C. de Barros}
 
\affil[1]{Instituto de Pesquisas Eldorado, Porto Alegre, RS, Brazil}
\affil[2]{Departamento de Física, Universidade Federal de Santa Catarina, Florianópolis, SC, Brazil}
\affil[3]{Instituto de Física Enrique Gaviola, Universidad Nacional de Córdoba, Argentina}
\affil[4]{Instituto de Informática, Universidade Federal do Rio Grande do Sul, Porto Alegre, RS, Brazil}

\date{}

\begin{document}

\maketitle


\begin{abstract}
We introduce a sparse Walsh-Pauli correlation encoding for MaxCut, in which graph variables are represented by diagonal Pauli-string expectation values computed classically as normalized XOR autocorrelations of Walsh coefficients. The complete Walsh basis can reproduce the diagonal correlators of arbitrary quantum states, while a structured sparse support reduces the number of trainable coefficients to $m+1$ for $m$ graph variables. We evaluate the method on $1{,}400$ held-out random graphs with up to $7{,}000$ vertices, obtaining empirical approximation ratios near $0.98$ relative to MQLib heuristic references. On Gset instances G14, G23, G60, and G64, the sparse model achieves mean refined approximation ratios of $0.9828$, $0.9926$, $0.9781$, and $0.9369$, respectively, with $80.4\%$-$89.3\%$ fewer parameters than the complete Walsh representation. The trained coefficients are transferred without retraining to quantum circuits. Statevector simulations of the four Gset instances achieve fidelities above $0.99994$ and preserve all correlator signs. Separately, a seven-qubit experiment on a $15$-vertex graph executed on IBM quantum hardware achieves a correlator-sign agreement of $0.927\pm0.049$ relative to the corresponding noiseless circuit. These results demonstrate the potential of sparse Walsh-PCE for large-instance quantum-inspired MaxCut optimization while validating the transfer of classically trained parameters to quantum hardware.
\end{abstract}

\section{Introduction}

Combinatorial optimization problems arise in network design, scheduling, logistics, statistical physics, and several other scientific and industrial domains. Many of these problems admit a quadratic unconstrained binary optimization (QUBO) formulation and can consequently be mapped to Ising-type objective functions~\citep{lucas2014}. MaxCut is a canonical example: given a weighted graph, the task is to partition its vertices into two subsets such that the total weight of the edges between the two subsets is maximized. Despite its simple formulation, MaxCut is NP-hard and remains an important benchmark for classical heuristics and quantum optimization algorithms~\citep{garey1976}.

Variational quantum algorithms (VQAs), including the Quantum Approximate Optimization Algorithm (QAOA), have been extensively investigated as hybrid approaches to combinatorial optimization~\citep{farhi2014qaoa,cerezo2021variational}. In the conventional encoding, each binary decision variable is associated with one qubit. Consequently, solving a graph problem with $m$ vertices requires at least $m$ qubits, in addition to a circuit sufficiently expressive to represent the relevant correlations. This one-to-one correspondence severely limits the size of the instances that can be addressed on current quantum processors.

Pauli-Correlation Encoding (PCE) provides an alternative representation in which decision variables are encoded in expectation values of Pauli strings rather than in individual qubits~\citep{sciorilli2025}. By associating each variable with a $k$-body Pauli observable acting on a smaller quantum register, PCE can represent a number of variables that grows polynomially with the number of qubits. This compression has enabled numerical and experimental studies of MaxCut instances substantially larger than the available quantum registers. Nevertheless, the original formulation still requires the optimization of a parametrized quantum circuit and repeated estimation of Pauli expectation values, making the training cost dependent on circuit simulation or quantum measurements.

In this work, we retain the qubit-efficient interpretation of PCE but restrict the encoded observables to mutually commuting products of Pauli-$Z$ operators. These observables are Walsh operators, whose eigenvalues in the computational basis are Walsh functions. Because Walsh functions form a complete basis for real-valued functions over binary domains, this correspondence provides a structured representation of diagonal Pauli correlations in terms of a real coefficient vector~\citep{walsh1923,welch2014}. The required expectation values can then be evaluated exactly as normalized XOR autocorrelations of these coefficients. The optimization stage can therefore be performed entirely in coefficient space, without repeatedly constructing a quantum circuit, sampling observables, or explicitly storing the corresponding state vector.

The Walsh representation also retains a direct circuit interpretation. Welch \emph{et al.} showed that Walsh operators form a basis for diagonal quantum operators and that diagonal unitaries expressed in this basis can be implemented using Pauli-$Z$ rotations and CNOT parity networks~\citep{welch2014}. Building on this construction, Zylberman and Debbasch introduced the Walsh Series Loader, which converts a Walsh-diagonal unitary into a function-encoded quantum state through ancilla-controlled interference and post-selection~\citep{zylberman2024}. Together, these results provide the circuit-level foundation used here: the Walsh coefficients are first trained classically and are subsequently frozen and inserted, without retraining, into a quantum circuit that reproduces the same encoded correlators.

The central remaining difficulty is the size of the Walsh representation. A complete expansion over $q$ qubits contains $2^q$ coefficients, while conventional truncation~\cite{welch2014, zylberman2024} may remove the XOR-related coefficient pairs required by the encoded correlators. To address this, we introduce a structured paired support that preserves these correlations. With a single base mask, the proposed construction requires only $m+1$ trainable parameters for $m$ graph variables, replacing the exponential Walsh representation with a parameterization that scales linearly with the problem size.

The contributions of this work are fourfold. First, we establish the connection between diagonal Pauli correlators and Walsh autocorrelations, showing that the complete Walsh basis can reproduce the diagonal correlators of arbitrary quantum states. Second, we introduce a structured paired support and investigate how its size affects solution quality. Third, we evaluate the complete and sparse formulations on $1{,}400$ held-out random graphs with up to $7{,}000$ vertices and on Gset instances~\cite{helmberg2000, ye_gset}. Fourth, we transfer classically trained coefficients to Walsh-based quantum circuits without retraining, validating the approach through statevector simulations and a separate experiment on superconducting quantum hardware.

A related independent approach by Lizzio Bosco et al.~\cite{lizzioBosco2026} also computes PCE correlators classically using efficiently simulable circuits. Our method instead uses closed-form normalized XOR autocorrelations and a structured paired support with $m+1$ parameters, while preserving direct transfer to quantum circuits.

The classical optimization and quantum-transfer stages serve distinct purposes. The quantum circuit is not used for training or improving MaxCut solutions, but to validate whether the classically optimized representation can be implemented and its correlators reconstructed from measurements. Thus, the hardware experiment demonstrates parameter transfer and measurability rather than quantum computational advantage.

This article is organized as follows. Sec.~\ref{sec:sparse_walsh_pce} introduces the diagonal PCE formulation, the Walsh representation of its correlators, and the structured paired-support construction. Sec.~\ref{sec:numerical_results} presents the numerical evaluation of the complete basis, the support-scaling analysis, the Gset benchmark, and the quantum-transfer experiments in simulation. Also, the real-device validation is reported in \cref{app:hardware_validation}. Finally, Sec.~\ref{sec:conclusion} summarizes the main findings, discusses the limitations of the present circuit implementation, and outlines directions for future work.

\section{Sparse Walsh-Pauli Correlation Encoding}
\label{sec:sparse_walsh_pce}
Consider a weighted graph $G=(V,E)$, where $V$ is the vertex set, $E$ is the edge set, and $w_{ij}$ is the weight associated with edge $(i,j)$. For each vertex $i$, the variable $z_i\in\{-1,+1\}$ indicates the part to which $i$ is assigned, while $\mathbf{z}$ denotes the assignment of all vertices. The MaxCut problem consists of partitioning the vertices into two sets so as to maximize the cut value, the total weight of the edges between the two sets, given by

\begin{equation}
    C(\mathbf{z})
    =
    \sum_{(i,j)\in E}
    w_{ij}
    \frac{1-z_i z_j}{2}.
    \label{eq:maxcut_objective}
\end{equation}

When $z_i$ and $z_j$ have opposite signs, edge $(i,j)$ crosses the cut and contributes $w_{ij}$; when their signs are equal, its contribution is zero. In quantum computing, MaxCut is commonly formulated as an Ising problem and addressed with variational quantum algorithms such as QAOA~\citep{farhi2014qaoa,lucas2014}. In the conventional encoding, each graph vertex is mapped directly to one qubit, and circuit measurement produces a bit string that defines the corresponding partition.

Although this representation is simple and natural, it imposes a one-to-one correspondence between variables and qubits, requiring $m$ qubits to represent a problem with $m$ vertices. This limitation motivates more compact encodings, such as Pauli-Correlation Encoding (PCE)~\citep{sciorilli2025}.

\subsection{Pauli-Correlation Encoding for MaxCut}
\label{sec:pce_maxcut}

In PCE, each binary variable is encoded by the sign of the expectation value of a Pauli observable~\citep{sciorilli2025}. In this work, to establish a direct connection with Walsh operators, we restrict the observable set to products of Pauli-$Z$ operators. For each vertex $i$, we therefore choose a distinct mask
\begin{equation}
    S_i\in\{0,1\}^{q},
    \qquad
    \lvert S_i\rvert=k,
\end{equation}
where $k$ is the product order, i.e., the number of Pauli-$Z$ factors in each observable, and $q$ is the number of qubits. The corresponding observable is
\begin{equation}
    \widehat W_{S_i}
    =
    \bigotimes_{r=1}^{q}
    \widehat Z_r^{(S_i)_r}.
    \label{eq:pce_diagonal_observable}
\end{equation}
For fixed $k$ and $q$, there are $\binom{q}{k}$ distinct weight-$k$ Pauli-$Z$ products. Hence, the encoding can represent $m\leq\binom{q}{k}$ vertices.

In the PCE formulation, for a $q$-qubit state
$\lvert\psi(\boldsymbol{\theta})\rangle$, the correlator representing
variable $i$ is
\begin{equation}
    c_i(\boldsymbol{\theta})
    =
    \langle\psi(\boldsymbol{\theta})\rvert
    \widehat W_{S_i}
    \lvert\psi(\boldsymbol{\theta})\rangle,
    \label{eq:pce_correlator}
\end{equation}
and the corresponding binary solution is obtained as
\begin{equation}
    z_i
    =
    \sign\left(c_i(\boldsymbol{\theta})\right)
    =
    \begin{cases}
        +1, & \text{if } c_i(\boldsymbol{\theta}) \geq 0,\\
        -1, & \text{if } c_i(\boldsymbol{\theta}) < 0.
    \end{cases}
    \label{eq:pce_binary_decoding}
\end{equation}
Here and throughout, we adopt the tie-breaking convention $\sign(0)=+1$. Because the sign function is not differentiable, during training it is replaced by the smooth relaxation
\begin{equation}
    s_i(\boldsymbol{\theta})
    =
    \tanh
    \left(
        \alpha c_i(\boldsymbol{\theta})
    \right),
    \label{eq:pce_relaxed_spin}
\end{equation}
where $\alpha>0$ sets the sharpness of the $\tanh$ response and restores a usable gradient. Following the PCE construction for MaxCut~\citep{sciorilli2025}, we use the relaxed objective

\begin{equation}
    \mathcal L(\boldsymbol{\theta})
    =
    \sum_{(i,j)\in E}
    w_{ij}
    s_i(\boldsymbol{\theta})
    s_j(\boldsymbol{\theta})
    +
    \beta \nu
    \left(
        \frac{1}{m}
        \sum_{i=1}^{m}
        s_i(\boldsymbol{\theta})^2
    \right)^2,
    \label{eq:pce_maxcut_loss}
\end{equation}
where $\beta>0$ controls the strength of the regularization and
the graph-dependent scale $\nu$ is defined as

\begin{equation}
    \nu
    =
    \frac{1}{2}
    \sum_{(i,j)\in E}
    \lvert w_{ij}\rvert
    +
    \frac{1}{4}
    W_F,
    \qquad
    W_F
    =
    \sum_{e\in F_{\min}}
    \lvert w_e\rvert,
    \label{eq:pce_loss_scales}
\end{equation}
where $F_{\min}$ is a minimum spanning forest of the graph
constructed using $\lvert w_{ij}\rvert$ as edge weights.
For a connected graph, $F_{\min}$ reduces to a minimum
spanning tree. In particular, for a connected unit-weight graph,

\begin{equation}
    \nu
    =
    \frac{|E|}{2}
    +
    \frac{m-1}{4}.
    \label{eq:pce_loss_scale_unit}
\end{equation}

The regularization penalizes large magnitudes of the relaxed
spins and counteracts their premature saturation near
$\pm1$, where the gradient of the $\tanh$ relaxation becomes
small. The parameter $\alpha$ controls the sharpness of the
relaxation, whereas $\beta$ controls this regularization
strength. Their numerical calibration is described in
\cref{sec:numerical_results}.
\subsection{Walsh Functions, Operators, and Function States}
\label{sec:walsh_function_states}

The Paley-ordered Walsh functions are defined for
$x\in[0,1)$ as~\citep{walsh1923,welch2014}
\begin{equation}
    w_j(x)
    =
    (-1)^{
        \sum_{r=1}^{q} j_r x_r
    },
    \label{eq:walsh_function}
\end{equation}
where $j_r\in\{0,1\}$ is the $r$-th bit in the binary expansion of
$j\in\{0,\ldots,2^q-1\}$, with
$j=\sum_{r=1}^{q}j_r2^{r-1}$. Moreover,
$x_r\in\{0,1\}$ is the $r$-th bit in the dyadic expansion of $x$,
$x=\sum_{r=1}^{\infty}x_r2^{-r}$.

For the discrete formulation, $x$ is sampled as
$x_k=k/N$, with $N=2^q$ and $k=0,\ldots,N-1$. We write
$k=\sum_{r=1}^{q}k_r2^{q-r}$, so that
$x_k=\sum_{r=1}^{q}k_r2^{-r}=0.k_1k_2\ldots k_q$. Defining
$w_{jk}=w_j(x_k)$, the discrete Walsh function can be written as
\begin{equation}
    w_{jk}
    =
    (-1)^{
        \sum_{r=1}^{q}j_rk_r
    }
    =
    (-1)^{
        \sum_{r=1}^{q}j_rx_r
    },
    \label{eq:walsh_binary_form}
\end{equation}
where $k_r$ is the $r$-th bit of $k$, ordered from the most significant to the least significant bit, equivalently the $r$-th digit in the dyadic expansion of $x_k$.

The complete set of $2^q$ Walsh functions has two properties central to the proposed construction. First, Walsh functions form an orthogonal basis for real functions defined on the $2^q$ discrete points, thus satisfying the orthogonality relation
\begin{equation}
    \sum_{k=0}^{2^q-1}
    w_{jk}w_{\ell k}
    =
    2^q\delta_{j\ell}.
    \label{eq:walsh_orthogonality}
\end{equation}
They also obey the product rule
\begin{equation}
    w_{jk}w_{\ell k}
    =
    w_{j\oplus\ell,k},
    \label{eq:walsh_xor_product}
\end{equation}
where $\oplus$ denotes bitwise XOR~\citep{walsh1923,welch2014}. Consequently, every real function defined on these points admits the expansion
\begin{equation}
    f_{\mathbf a}(x_k)
    =
    \sum_{\ell=0}^{2^q-1}
    a_\ell w_{\ell k},
    \label{eq:walsh_expansion}
\end{equation}
with coefficients obtained from the orthogonality relation in \cref{eq:walsh_orthogonality} as
\begin{equation}
    a_\ell
    =
    \frac{1}{2^q}
    \sum_{k=0}^{2^q-1}
    f_{\mathbf a}(x_k)w_{\ell k}.
    \label{eq:walsh_coefficients}
\end{equation}

The expansion in \cref{eq:walsh_expansion} can be used to construct a quantum state $\ket{f}$. For this, we use the Walsh operator defined by~\cite{welch2014, zylberman2024}
\begin{equation}
    \widehat W_j
    =
    \bigotimes_{r=1}^{q}
    (\widehat Z_r)^{j_r},
    \label{eq:walsh_operator}
\end{equation}
which is a Pauli-Z product, the same as that used in the PCE in \cref{eq:pce_diagonal_observable}. Since
\begin{equation}
    \widehat Z_r
    \lvert k_r\rangle
    =
    (-1)^{k_r}
    \lvert k_r\rangle,
    \label{eq:pauli_z_action}
\end{equation}
the action of $\widehat W_j$ on
$\lvert k\rangle=\lvert k_1,\ldots,k_q\rangle$ is
\begin{equation}
    \widehat W_j
    \lvert k\rangle
    =
    (-1)^{
        \sum_{r=1}^{q}j_rk_r
    }
    \lvert k\rangle
    =
    w_{jk}\lvert k\rangle,
    \label{eq:walsh_operator_action}
\end{equation}
in agreement with the discrete definition in \cref{eq:walsh_binary_form}. Thus, Walsh operators are tensor products of $\mathbb I$ and Pauli-$Z$, and Walsh functions provide their eigenvalues in the computational basis~\citep{welch2014}.

The coefficients in \cref{eq:walsh_expansion} define the diagonal operator
\begin{equation}
    \widehat F(\mathbf a)
    =
    \sum_{\ell=0}^{2^q-1}
    a_\ell\widehat W_\ell.
    \label{eq:walsh_function_operator}
\end{equation}
Applying this operator and using \cref{eq:walsh_expansion} and  yields \cref{eq:walsh_operator_action}
\begin{equation}
    \widehat F(\mathbf a)
    \lvert+\rangle^{\otimes q}
    =
    \frac{1}{\sqrt{2^q}}
    \sum_{k=0}^{2^q-1}
    f_{\mathbf a}(x_k)
    \lvert k\rangle.
    \label{eq:walsh_operator_generates_function}
\end{equation}
For any nonzero coefficient vector $\mathbf a$, normalizing the state in \cref{eq:walsh_operator_generates_function} gives
\begin{equation}
    \lvert f(\mathbf a)\rangle
    =
    \frac{1}{\norm{f_{\mathbf a}}_2}
    \sum_{k=0}^{2^q-1}
    f_{\mathbf a}(x_k)
    \lvert k\rangle,
    \label{eq:walsh_function_state}
\end{equation}
where
$\norm{f_{\mathbf a}}_2^2
=\sum_{k=0}^{2^q-1}f_{\mathbf a}(x_k)^2$.

Therefore, we can use the Walsh series to construct any state with real computational-basis amplitudes. Furthermore, we note that the Walsh operators are the very terms used in the PCE formalism, allowing us to use the state $\ket{f}$, along with the properties of Walsh functions, to produce dense and sparse representations.

\subsection{Walsh Representation of Diagonal PCE Correlations}
\label{sec:walsh_pce_connection}
For any mask $S$, using the Walsh product rule in
\cref{eq:walsh_xor_product} and the orthogonality relation in \cref{eq:walsh_orthogonality}, the expectation value of the corresponding Walsh operator in the state $\ket{f(\mathbf a)}$ can be
written as
\begin{equation}
    c_S(\mathbf a)
    =
    \frac{
        \sum_{\ell=0}^{2^q-1}
        a_\ell
        a_{\ell\oplus S}
    }{
        \sum_{\ell=0}^{2^q-1}
        a_\ell^2
    },
    \label{eq:walsh_autocorrelation}
\end{equation}
establishing the exact connection between the two formalisms. Although each encoded variable remains the expectation value of a Pauli-$Z$ product, it can now be evaluated directly as a normalized XOR autocorrelation of the Walsh coefficients. The trainable parameters are therefore exclusively
$\mathbf a=(a_0,\ldots,a_{2^q-1})$.

This formulation allows all relevant quantities to be evaluated directly in Walsh-coefficient space, without explicitly constructing either the quantum circuit or the full state vector. At the level of diagonal observables, the complete Walsh basis is fully expressive: whenever an optimal diagonal-PCE solution exists, it can reproduce the same correlators, objective value, and decoded solution, as shown in \cref{app:walsh_representability}. At the same time, the coefficients retain a direct operational meaning and can subsequently be mapped to a quantum circuit, as discussed in \cref{sec:gate_implementation}.

The main limitation of the complete representation is its exponential size, since it requires $2^q$ coefficients. Although truncated Walsh expansions can approximate diagonal operators with fewer terms~\citep{welch2014}, a standard coefficient-wise truncation is not appropriate in the present setting. Each PCE correlator depends on pairs of coefficients whose indices differ by a prescribed XOR mask. Removing coefficients independently may therefore eliminate the pairs required to represent specific variables and force the corresponding correlators to vanish. Any support-reduction strategy must consequently preserve this XOR-pair structure.

\subsection{Structured Paired Support}
\label{sec:structured_paired_support}
To obtain a sparse parameterization without eliminating the correlators required by PCE, the coefficient support must be constructed structurally. In particular, for every mask associated with a problem variable, pairs of indices connected by XOR must be preserved. This condition motivates a paired active support.

For an active support
$\mathcal A\subseteq\{0,\ldots,2^q-1\}$, define
\begin{equation}
    f_{\mathcal A}(\mathbf x)
    =
    \sum_{\ell\in\mathcal A}
    a_\ell
    w_\ell(\mathbf x).
    \label{eq:sparse_walsh_function}
\end{equation}
Then, \cref{eq:walsh_autocorrelation} becomes
\begin{equation}
    c_S^{(\mathcal A)}(\mathbf a)
    =
    \frac{
        \sum_{\substack{
            \ell\in\mathcal A\\
            \ell\oplus S\in\mathcal A
        }}
        a_\ell
        a_{\ell\oplus S}
    }{
        \sum_{\ell\in\mathcal A}
        a_\ell^2
    }.
    \label{eq:sparse_walsh_autocorrelation}
\end{equation}
Thus, $c_S$ receives contributions only from active index pairs related by $S$ through XOR. To guarantee the presence of such pairs, let
$\mathcal S=\{S_i\}_{i=1}^{m}$ denote the decision-mask set and
$\mathcal B=\{b_r\}_{r=1}^{B}$ the base-mask set, where $B=\lvert\mathcal B\rvert$.

Because the correlator associated with $S_i$ receives contributions only from index pairs separated by that mask, we construct the support from the base-mask set $\mathcal B=\{b_r\}_{r=1}^{B}$. For every decision mask $S_i\in\mathcal S$, the support includes both $b_r$ and the shifted index $b_r\oplus S_i$. We therefore define
\begin{equation}
    \mathcal A(\mathcal B,\mathcal S)
    =
    \mathcal B
    \cup
    \left\{
        b_r\oplus S_i:
        b_r\in\mathcal B,\;
        S_i\in\mathcal S
    \right\}.
    \label{eq:structured_paired_support}
\end{equation}

The set $\mathcal B$ contains only the indices used to generate the active support and it does not define a different functional basis. By construction, for every $S_i\in\mathcal S$ and $b_r\in\mathcal B$, both $b_r$ and $b_r\oplus S_i$ belong to $\mathcal A$. Hence, even before training, the numerator of \cref{eq:sparse_walsh_autocorrelation} contains the pair $a_{b_r}a_{b_r\oplus S_i}$. This prevents the correlator associated with $S_i$ from being structurally zero solely because compatible indices are absent from the support. Already for $B=1$, i.e.\ the single-base support $\mathcal A=\{0\}\cup\mathcal S$, sign universality holds: every binary assignment is the decoded solution of some coefficient vector (\cref{prop:sign_universality}), and larger supports inherit this. Representability of binary solutions is therefore not a limitation of the single-base model.

Because each base mask can generate at most one original index and $m$ shifted indices, the
number of active coefficients satisfies
\begin{equation}
    |A(B,\mathcal S)| \leq B(m+1).
\end{equation}
This bound characterizes the number of trainable coefficients, rather than the total computational cost, which also depends on the number of XOR-compatible coefficient pairs and on the graph edges entering the loss.
The actual number can be smaller when different combinations of $b_r$ and $S_i$ produce the same index under XOR. For clarity, an explicit example of the correlator construction and the distinction between the Walsh basis, the base masks, and the active support is given in \cref{app:structured_support_example}.

The resulting formulation defines a hierarchy of Walsh-PCE models controlled by the base-mask set $\mathcal B$. The complete Walsh basis provides the maximally expressive diagonal representation, whereas smaller values of $B$ progressively restrict the number of coefficient pairs contributing to each decision correlator. The paired-support construction guarantees that every encoded variable retains at least one admissible XOR-related pair, but it does not, by itself, guarantee that a reduced support preserves the solution quality of the complete representation. The numerical analysis therefore proceeds in three stages: we first establish the performance of the complete basis, then quantify the effect of reducing and scaling the paired support, and finally evaluate the selected sparse model on public benchmarks and through quantum-circuit transfer.

\section{Numerical Results}
\label{sec:numerical_results}

We now evaluate the Walsh-PCE formulation introduced in the previous
sections. The numerical analysis proceeds in three stages. We first use
the complete Walsh basis as the most expressive diagonal reference,
then investigate how far the active support can be reduced while
preserving solution quality, and finally evaluate the resulting sparse
model on public Gset instances. The trained sparse coefficients are
subsequently transferred to the quantum construction described in
\cref{sec:gate_implementation}.

Throughout the classical experiments, the trainable parameters are the
active Walsh coefficients entering the autocorrelation formula in
\cref{eq:walsh_autocorrelation}. The corresponding correlators are
mapped to the relaxed variables defined in
\cref{eq:pce_relaxed_spin} and optimized through the MaxCut loss in
\cref{eq:pce_maxcut_loss}. Unless otherwise stated, optimization is
performed in double precision using Adam~\citep{kingma2014adam}, with
learning rate $\eta=10^{-3}$ for $500$ epochs and ten independent
coefficient initializations. Gradients are evaluated by automatic
differentiation over all graph edges, with no early stopping.

For each initialization, we set $a_0=1$ and sample the remaining coefficients from $\mathcal N(0,10^{-4})$. The coefficient vector is normalized to unit Euclidean norm initially and after each Adam update, since the correlators in~\cref{eq:walsh_autocorrelation} are invariant under global rescaling.

When different active supports are compared, a common initialization is first generated in global Walsh-index order and then restricted to the coefficients belonging to each support. The resulting restricted vector is subsequently renormalized. Thus, models compared under the same initialization inherit compatible coefficient values rather than independently generated random vectors.

Solution quality is reported through the empirical approximation ratio
\begin{equation}
    \mathrm{AR}(G)
    =
    \frac{C_{\mathrm{found}}(G)}
         {C_{\mathrm{ref}}(G)},
    \label{eq:empirical_ar}
\end{equation}
where $C_{\mathrm{found}}(G)$ is the cut returned by the Walsh-PCE
solution and $C_{\mathrm{ref}}(G)$ is an external reference value.
For the random-graph experiments, the reference is obtained with the MQLib hyper-heuristic~\citep{dunning2018}. Each graph is evaluated over ten independent MQLib runs with a nominal runtime budget of $5\,\mathrm{s}$ per run, and the best observed cut is retained. Since this is a heuristic reference rather than a certified optimum, empirical approximation ratios slightly larger
than one may occur. MQLib is used only to normalize solution quality and not as a direct runtime baseline. For the Gset experiments, $C_{\mathrm{ref}}$ is instead taken from the corresponding published best-known values, as described in \cref{sec:gset_benchmark}.

Unless explicitly indicated otherwise, the reported Walsh-PCE solutions are evaluated directly after decoding the optimized correlators, without local-search refinement. The bit-flip post-processing step is considered separately in the Gset analysis.

\subsection{Performance of the Complete Basis on Random Graphs}
\label{sec:full_basis_random_graphs}

We first evaluate the complete Walsh representation. As established by
the representability result in \cref{app:walsh_representability}, the
complete basis can reproduce any solution characterized by the diagonal
Pauli-$Z$ correlators used by the PCE formulation. Its numerical
performance is not interpreted as a certified upper bound, since the
optimized solution still depends on the variational landscape and
training dynamics. Instead, it provides the most expressive diagonal
reference for the support-reduction analysis.

We considered unweighted Erd\H{o}s--R\'enyi graphs with
$m\in\{100,200,400,500,800,2000,7000\}$ vertices. For each size,
instances were generated from \(G(m,p)\)~\citep{erdos1959}, with
\(p=6/(m-1)\), corresponding to a target mean degree of six. Graphs with
mean degree below three were discarded, while disconnected accepted
instances were made connected by adding the minimum number of
unit-weight edges between components.

For each graph size, \(250\) instances were retained, yielding
\(1{,}750\) graphs in total. Of these, \(50\) instances per size were
reserved exclusively for hyperparameter calibration and the remaining
\(200\) for held-out evaluation, resulting in \(350\) calibration and
\(1{,}400\) evaluation graphs. The same held-out instances are reused
in the support-reduction analysis of \cref{sec:base_mask_scaling}.

The decision-mask construction follows \cref{sec:pce_maxcut}. For each
graph size, $q$ is chosen as the smallest register size satisfying
\(\binom{q}{k}\geq m\). When more than $m$ Hamming-weight-$k$ masks are available, the first $m$ masks in lexicographic order of their qubit supports are assigned to the vertices. The resulting $(q,k)$ configurations are reported in \cref{tab:full_basis_performance}.

\begin{table}[!h]
\centering
\caption{Performance of the complete Walsh basis on held-out random graphs. For each graph size, results cover $200$ graphs and ten initializations ($2{,}000$ runs). Standard deviation and $p_5$ are computed over all runs, while \emph{best-of-10} is averaged across graphs.
}
\label{tab:full_basis_performance}
\begin{tabular}{rrrcccc}
\toprule
$m$
& $q$
& $k$
& $\overline{\mathrm{AR}}\pm\sigma$
& $p_{5}$
& \emph{best-of-10}
& $\mathrm{AR}\geq0.98$ \\
\midrule
$100$     & $9$  & $4$ & $0.9762 \pm 0.0135$ & $0.9515$ & $0.9923$ & $0.419$ \\
$200$     & $10$ & $5$ & $0.9759 \pm 0.0099$ & $0.9583$ & $0.9892$ & $0.364$ \\
$400$     & $11$ & $5$ & $0.9765 \pm 0.0072$ & $0.9649$ & $0.9867$ & $0.318$ \\
$500$     & $12$ & $6$ & $0.9752 \pm 0.0065$ & $0.9643$ & $0.9843$ & $0.241$ \\
$800$     & $12$ & $6$ & $0.9826 \pm 0.0054$ & $0.9735$ & $0.9900$ & $0.695$ \\
$2{,}000$ & $14$ & $7$ & $0.9792 \pm 0.0033$ & $0.9737$ & $0.9841$ & $0.403$ \\
$7{,}000$ & $16$ & $8$ & $0.9860 \pm 0.0066$ & $0.9793$ & $0.9887$ & $0.894$ \\
\bottomrule
\end{tabular}
\end{table}

The hyperparameters $\alpha$ and $\beta$, defined in
\cref{eq:pce_relaxed_spin,eq:pce_maxcut_loss}, were selected using only
the calibration subset. Calibration was performed with the complete
Walsh basis independently for each encoding configuration $(q,k)$,
rather than independently for each graph size. Consequently, the
$m=500$ and $m=800$ instances, which share $(q,k)=(12,6)$, were pooled
during calibration.

Calibration was performed sequentially. First, $\alpha$ was selected
with $\beta=0.5$ fixed; the selected $\alpha$ was then held fixed while
$\beta$ was varied. Each candidate was evaluated over five independent
coefficient initializations using Adam with learning rate
$\eta=10^{-3}$ for $150$ epochs. Candidates were ranked by the mean
empirical approximation ratio over all calibration graphs and
initializations belonging to the corresponding encoding configuration.
In the event of an exact tie, the smaller candidate value was selected.
No held-out graph and no Gset instance was used during calibration.

The selected hyperparameters are reported in~\cref{tab:hyperparameters_random_graphs}. The selected values were then fixed for all subsequent complete-basis and support-reduction experiments using the corresponding encoding configuration. The final complete-basis evaluation uses the $200$ held-out graphs of each size and ten independent initializations per graph.
\begin{table}[!h]
\centering
\caption{Hyperparameters selected from the disjoint calibration subset
for each encoding configuration.}
\label{tab:hyperparameters_random_graphs}
\begin{tabular}{rrr}
\toprule
$(q,k)$ & $\alpha$ & $\beta$ \\
\midrule
$(9,4)$   & $27$ & $0.75$ \\
$(10,5)$  & $22$ & $0.75$ \\
$(11,5)$  & $22$ & $0.75$ \\
$(12,6)$  & $36$ & $0.75$ \\
$(14,7)$  & $42$ & $0.50$ \\
$(16,8)$  & $80$ & $0.50$ \\
\bottomrule
\end{tabular}
\end{table}

As shown in \cref{tab:full_basis_performance}, the complete basis consistently produces high-quality solutions across all evaluated graph sizes, with mean empirical approximation ratios between $0.9752$ and $0.9860$. The fifth percentile indicates strong performance even in the lower tail, while the \emph{best-of-10} statistic captures the benefit of multiple independent initializations. The reported standard deviation reflects variability across both graph instances and initializations and should therefore not be interpreted solely as sensitivity to initialization.

\subsection{Support Reduction and Base-Mask Scaling}
\label{sec:base_mask_scaling}

Having established the complete basis as the reference representation,
we next investigate how strongly the paired Walsh support can be reduced
before solution quality deteriorates. Following
\cref{eq:structured_paired_support}, a base-mask set
$\mathcal B$ with $B=\lvert\mathcal B\rvert$ generates the active
support. The most restrictive case uses a single base mask,
\begin{equation}
    \mathcal B=\{0\},
    \qquad
    \mathcal A=\{0\}\cup\mathcal S,
    \qquad
    \lvert\mathcal A\rvert=m+1.
    \label{eq:minimal_support}
\end{equation}

The support-scaling study uses the same $200$ held-out graphs per size, with five independent initializations for each support size $B$. All runs follow the optimization procedure described above.

For each encoding configuration, we generate a fixed ordering of candidate base masks. The first $B$ masks define $\mathcal B$, so increasing $B$ only adds masks to the existing set. This ensures that the support selection is independent of the coefficient initialization. Because different XOR combinations may produce the same Walsh index, the actual support size can be smaller than $B(m+1)$. The base masks and resulting support sizes are reported in \cref{tab:support_grid}.

\begin{table}[!h]
\centering
\caption{Base-mask values $B$ and resulting active-support sizes
$\lvert\mathcal A\rvert$ used in the support-scaling experiment.
The final entry for each graph size is the first value
$B_{\mathrm{sat}}$ for which the paired support reaches the complete
Walsh basis.}
\label{tab:support_grid}
\setlength{\tabcolsep}{4pt}
\resizebox{\linewidth}{!}{%
\begin{tabular}{rll}
\toprule
$m$ & $B$ & resulting $\lvert\mathcal A\rvert$ \\
\midrule
$100$
& $1,2,4,8,16,23$
& $101,202,324,409,494,512$ \\

$200$
& $1,2,4,8,16,32,33$
& $201,400,616,798,983,1021,1024$ \\

$400$
& $1,2,4,8,16,32,34$
& $401,802,1271,1727,1986,2046,2048$ \\

$500$
& $1,2,4,8,16,32,61$
& $501,792,1407,2699,3592,4070,4096$ \\

$800$
& $1,2,4,8,16,32,37$
& $801,1602,2556,3556,4011,4095,4096$ \\

$2{,}000$
& $1,2,4,8,16,32,64,89$
& $2001,3834,6889,10042,14607,16211,16380,16384$ \\

$7{,}000$
& $1,2,4,8,16,32,64,85$
& $7001,14002,23104,41981,55126,63676,65521,65536$ \\
\bottomrule
\end{tabular}%
}
\end{table}

For each graph size, the final value $B_{\mathrm{sat}}$ in
\cref{tab:support_grid} is the first base-mask count for which the
paired support reaches all $2^q$ Walsh coefficients. This saturation
point is not retrained as an additional sparse model; instead, the
corresponding complete-basis result from
\cref{sec:full_basis_random_graphs} is used as the endpoint of the
support sweep.

In addition to the empirical approximation ratio in \cref{eq:empirical_ar}, sparse solutions are compared with the complete-basis solutions using the recovery ratio
\begin{equation}
    R
    =
    \frac{C(\mathbf z_B)}
         {C(\mathbf z_{\mathrm{full}})},
    \label{eq:support_recovery}
\end{equation}
where both cut values correspond to the same graph and initialization. Recovery statistics are aggregated over the five paired initializations per graph.

\begin{figure}[!h]
\centering
\includegraphics[width=0.85\linewidth]{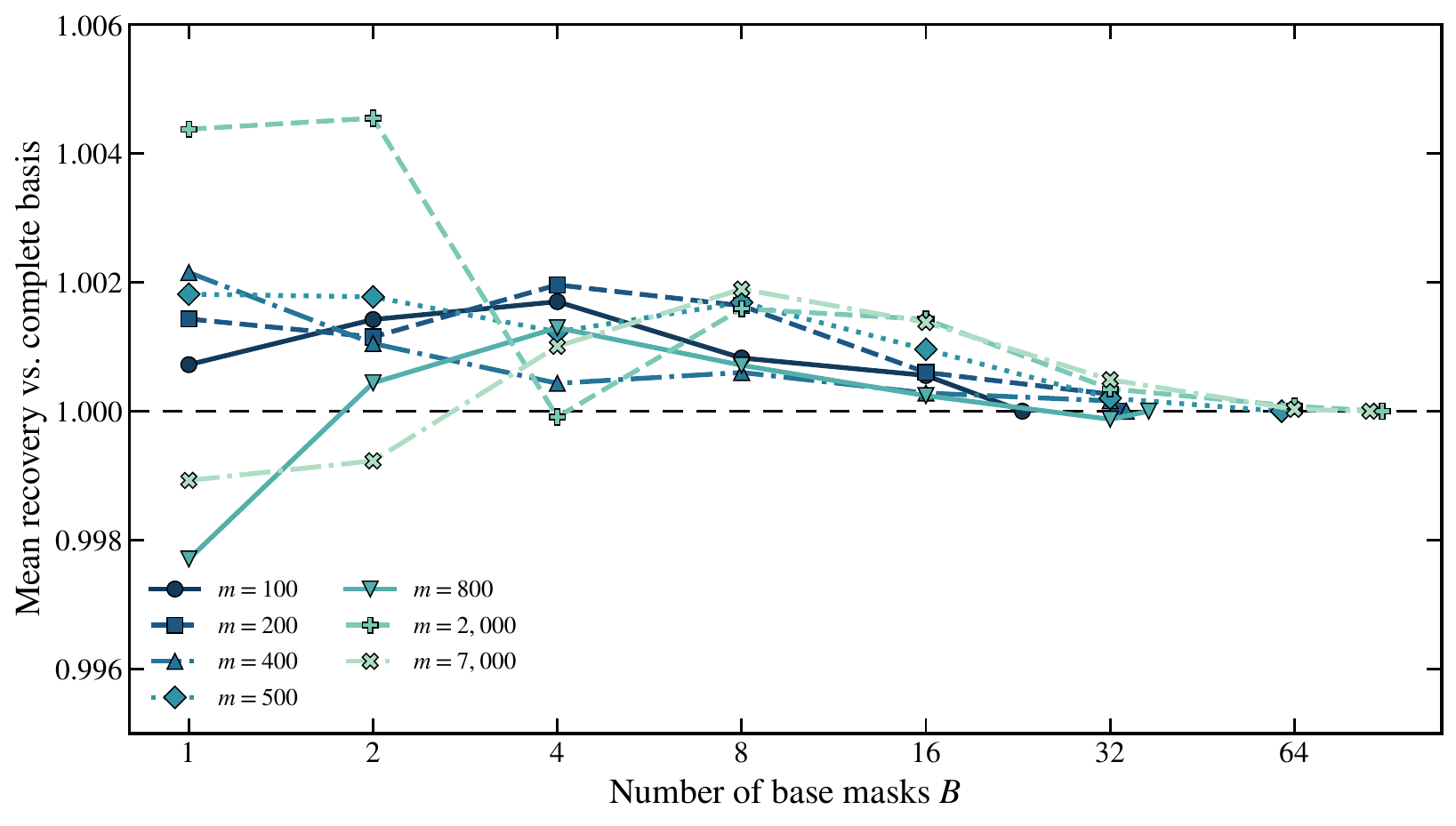}
\caption{Base-mask scaling on the held-out random-graph evaluation set. Sparse points use $200$ graphs per size and five coefficient initializations. Recovery is measured relative to the complete-basis solution for the same graph and initialization. The largest $B$ shown for each graph size corresponds to the complete-basis reference imported from \cref{sec:full_basis_random_graphs}, rather than to a separately retrained sparse model.}
\label{fig:base_mask_sweep}
\end{figure}

\cref{fig:base_mask_sweep} summarizes the resulting support-scaling behavior. Increasing $B$ does not produce a systematic monotonic improvement in solution quality. Recovery remains close to unity across the evaluated support sizes, including the single-base model, with no zero-cut collapse observed.

The single-base model already achieves solution quality comparable to the complete basis while using only $m+1$ trainable coefficients. Its performance is summarized in \cref{tab:base_mask_b1}.

\begin{table}[!h]
\centering
\caption{Single-base model ($B=1$, $\lvert\mathcal A\rvert=m+1$) on the held-out random-graph evaluation set. The retained fraction is measured relative to the complete Walsh basis, and recovery is measured relative to the complete-basis solution for the same graph and initialization.}
\label{tab:base_mask_b1}
\begin{tabular}{rrrrcc}
\toprule
$m$
& $\lvert\mathcal A\rvert$
& retained
& $\mathrm{AR}$
& recovery
& $p_5$ recovery \\
\midrule
$100$     & $101$     & $19.73\%$ & $0.9770 \pm 0.0136$ & $1.0007 \pm 0.0130$ & $0.9789$ \\
$200$     & $201$     & $19.63\%$ & $0.9769 \pm 0.0100$ & $1.0014 \pm 0.0108$ & $0.9826$ \\
$400$     & $401$     & $19.58\%$ & $0.9787 \pm 0.0071$ & $1.0022 \pm 0.0084$ & $0.9890$ \\
$500$     & $501$     & $12.23\%$ & $0.9771 \pm 0.0066$ & $1.0018 \pm 0.0076$ & $0.9892$ \\
$800$     & $801$     & $19.56\%$ & $0.9804 \pm 0.0053$ & $0.9977 \pm 0.0066$ & $0.9876$ \\
$2{,}000$ & $2{,}001$ & $12.21\%$ & $0.9835 \pm 0.0031$ & $1.0044 \pm 0.0042$ & $0.9975$ \\
$7{,}000$ & $7{,}001$ & $10.68\%$ & $0.9850 \pm 0.0066$ & $0.9989 \pm 0.0025$ & $0.9949$ \\
\bottomrule
\end{tabular}
\end{table}

The single-base support retains only $10.7\%$ to $19.7\%$ of the coefficients of the corresponding complete Walsh representation, reducing the number of trainable parameters by approximately $80.3\%$ to $89.3\%$ while preserving comparable cut quality. In some cases, the mean recovery slightly exceeds unity. This does not indicate greater expressivity, since the sparse model is contained within the complete Walsh space. Rather, under a fixed optimization budget, the smaller parameter space may facilitate optimization, while stochastic variability can also contribute to these differences.

These results indicate that progressively increasing the support size is not necessary to achieve solution quality comparable to the complete basis. Under the fixed optimization budget considered here, the single-base support is already sufficient on the investigated random graphs. We therefore use $B=1$ as the sparse configuration in the remainder of the work.


\subsection{Gset Benchmark}
\label{sec:gset_benchmark}

To evaluate the single-base model on established benchmarks, we use the Gset collection~\citep{helmberg2000,ye_gset}. Unlike the random graphs considered above, Gset instances are fixed and publicly available, allowing comparisons across studies.

We select four instances with $m=800$ to $m=7000$ vertices, as summarized in \cref{tab:gset_instances}. G14, G23, and G60 have unit positive weights, whereas G64 has the same number of vertices as G60 but more than twice as many edges and signed $\pm1$ weights, providing a more challenging test case.

Solution quality is normalized by the adopted best-known cut value $C_{\mathrm{ref}}$~\citep{shylo2015}, which is a benchmark reference rather than a certified optimum, particularly for G60 and G64. G14, G23, and G64 are connected, whereas G60 has $45$ connected components. For G60, the regularization scale therefore follows the minimum-spanning-forest definition in \cref{eq:pce_loss_scales}.

\begin{table}[!h]
\centering
\caption{Gset instances used in the benchmark and the reference cut values $C_{\mathrm{ref}}$ adopted to normalize solution quality.}
\label{tab:gset_instances}
\begin{tabular}{lrrcr}
\toprule
Instance & $m$ & $\lvert E\rvert$ & weights & $C_{\mathrm{ref}}$ \\
\midrule
G14 & $800$     & $4694$  & $+1$   & $3064$  \\
G23 & $2{,}000$ & $19990$ & $+1$   & $13344$ \\
G60 & $7{,}000$ & $17148$ & $+1$   & $14188$ \\
G64 & $7{,}000$ & $41459$ & $\pm1$ & $8751$  \\
\bottomrule
\end{tabular}
\end{table}

Each instance is solved using the single-base model ($B=1$) with ten independent initializations, following the optimization protocol defined above. The values of $\alpha$ and $\beta$ are transferred directly from the corresponding random-graph calibration, without recalibration on Gset.

We report both the directly decoded solution (\emph{raw}) and the result after local refinement (\emph{refined}). The refinement consists of one forward greedy sweep over vertices $0,\ldots,m-1$, followed by one backward sweep over $m-1,\ldots,0$. A vertex is flipped only if it strictly increases the cut value, with no additional sweeps.

\begin{table}[!h]
\centering
\caption{Single-base ($B=1$) results on Gset. The table reports the encoding $(q,k)$, active coefficients $\lvert\mathcal A\rvert=m+1$ relative to the complete basis $2^q$, parameter reduction, and raw and refined empirical ARs (mean $\pm$ standard deviation and best value over ten initializations).}
\label{tab:gset_ours}
\setlength{\tabcolsep}{4pt}
\resizebox{\linewidth}{!}{%
\begin{tabular}{lccccccc}
\toprule
Instance & $(q,k)$ & $\lvert\mathcal A\rvert \,/\, 2^{q}$ & reduction
& raw AR & best & refined AR & best \\
\midrule
G14 & $(12,6)$ & $801 \,/\, 4{,}096$       & $80.4\%$ & $0.9698 \pm 0.0062$ & $0.9827$ & $0.9828 \pm 0.0029$ & $0.9879$ \\
G23 & $(14,7)$ & $2{,}001 \,/\, 16{,}384$  & $87.8\%$ & $0.9874 \pm 0.0016$ & $0.9898$ & $0.9926 \pm 0.0016$ & $0.9957$ \\
G60 & $(16,8)$ & $7{,}001 \,/\, 65{,}536$  & $89.3\%$ & $0.9746 \pm 0.0015$ & $0.9772$ & $0.9781 \pm 0.0016$ & $0.9808$ \\
G64 & $(16,8)$ & $7{,}001 \,/\, 65{,}536$  & $89.3\%$ & $0.9195 \pm 0.0060$ & $0.9330$ & $0.9369 \pm 0.0058$ & $0.9491$ \\
\bottomrule
\end{tabular}%
}
\end{table}

As shown in \cref{tab:gset_ours}, the single-base model achieves high approximation ratios on G14, G23, and G60, with parameter reductions of $80.4\%$ to $89.3\%$ relative to the complete Walsh basis. The refined mean ARs range from $0.9781$ to $0.9926$, with relatively small improvements from local search. This indicates that the Walsh-PCE correlator signs already produce high-quality partitions before post-processing.

G64 presents the lowest approximation ratios, with a mean raw AR of $0.9195$ and refined AR of $0.9369$. Its larger local-search gain suggests greater dependence on classical post-processing. However, because G64 differs in both graph density and edge-sign structure, these results do not isolate the cause of the lower performance. We therefore interpret this as an empirical limitation of the present configuration rather than an intrinsic limitation of the single-base representation.

Overall, the Gset results extend the findings from random graphs to established benchmark instances, showing that the single-base model maintains high approximation ratios despite substantial parameter reduction. Although performance varies across graph structures, these results support the use of $B=1$ as a compact representation for large-scale MaxCut optimization.


\subsection{Quantum Transfer in Simulation}
\label{sec:quantum_transfer_simulation}

Once trained, the coefficient vector $\mathbf a^\star$ of the single-base model defines a diagonal Walsh operator and its corresponding transfer unitary in \cref{eq:transfer_unitary}. Following the Walsh-based circuit construction described in \cref{sec:gate_implementation}, the trained coefficients are transferred without additional optimization. For each Gset instance, we select the initialization with the highest raw empirical AR, freeze its coefficients, and prepare the corresponding state through ancilla interference and post-selection, as defined in \cref{eq:finite_epsilon_state}.

To evaluate the transfer, we compare the target Walsh function state $\ket{f(\mathbf a^\star)}$ with the post-selected state $\ket{\phi_\epsilon(\mathbf a^\star)}$ obtained from statevector simulation. Their agreement is quantified by the pure-state fidelity,
\begin{equation}
F
=
\left|
\left\langle
f(\mathbf a^\star)
\middle|
\phi_{\epsilon}(\mathbf a^\star)
\right\rangle
\right|^{2},
\label{eq:transfer_fidelity}
\end{equation}
and the total-variation (TV) distance between their computational-basis probability distributions,
\begin{equation}
D_{\mathrm{TV}}(p,q)
=
\frac{1}{2}
\sum_x
\left|
p(x)-q(x)
\right|,
\label{eq:tv_distance}
\end{equation}
where $p(x)=|\langle x|f(\mathbf a^\star)\rangle|^2$ and $q(x)=|\langle x|\phi_\epsilon(\mathbf a^\star)\rangle|^2$. Fidelity measures agreement between the complete quantum states, whereas TV distance compares their measurement distributions.

Since the decision observables are diagonal, their correlators depend only on these distributions. Using the Walsh eigenvalues defined in \cref{eq:pce_diagonal_observable}, the difference between correlators satisfies
\begin{equation}
\left|c_S(p)-c_S(q)\right|
\leq
2D_{\mathrm{TV}}(p,q).
\label{eq:corr_tv_bound}
\end{equation}
Thus, the TV distance bounds the error in every decision correlator, guaranteeing sign preservation whenever $|c_S(p)|>2D_{\mathrm{TV}}(p,q)$. We also evaluate the mean absolute error (MAE), maximum absolute error, and fraction of preserved correlator signs, which determine whether the decoded MaxCut solution changes.

\begin{figure}[H]
\centering
\includegraphics[width=\linewidth]{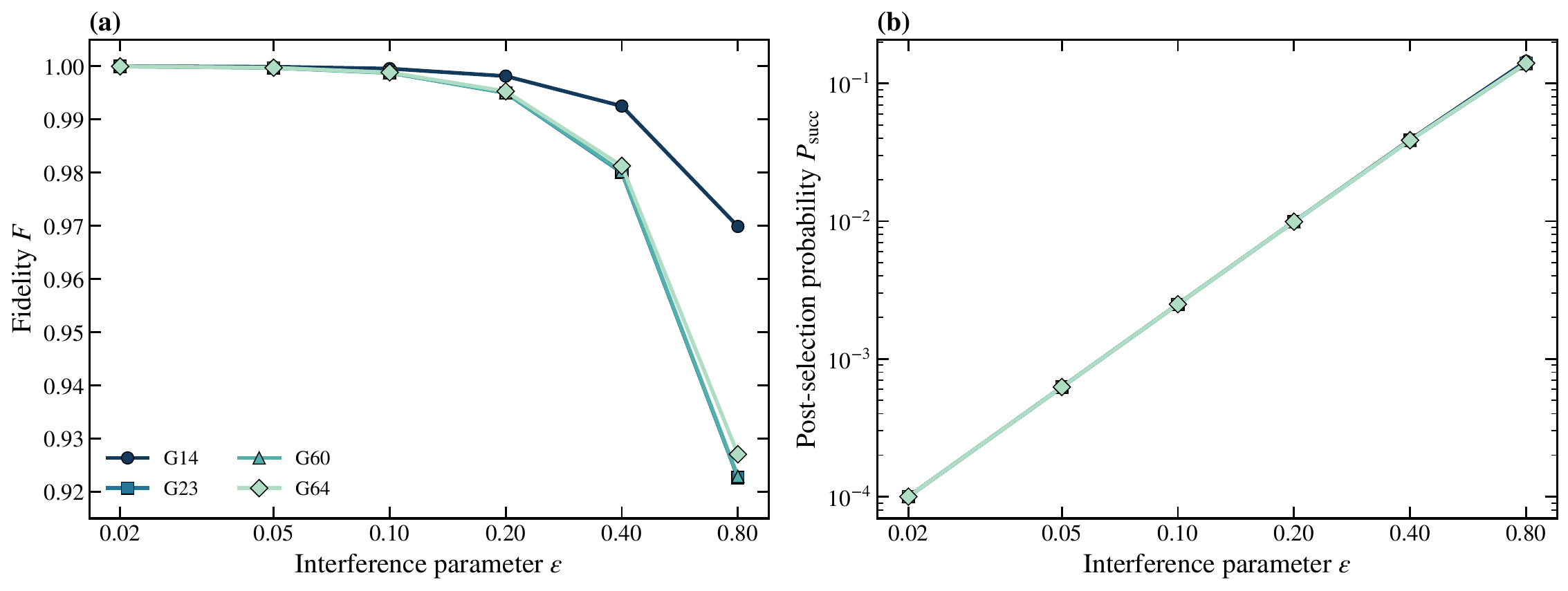}
\caption{Dependence of (a) state fidelity and (b) post-selection success probability on $\epsilon$ for G14, G23, G60, and G64. Smaller $\epsilon$ improves fidelity at the cost of lower post-selection probability.}
\label{fig:epsilon_window}
\end{figure}

The transfer accuracy depends on the interference parameter $\epsilon$, which controls the trade-off between state fidelity and post-selection success probability~\citep{zylberman2024}. As shown in \cref{fig:epsilon_window}, smaller values improve fidelity but reduce the probability of successful state preparation. For the statevector validation, we select $\epsilon=0.02$, the smallest tested value, to minimize finite-angle distortion.

The statevector results at $\epsilon=0.02$ are summarized in \cref{tab:transfer_statevector}. For all four instances, fidelity exceeds $0.99994$, TV distance remains of order $10^{-5}$, and correlator errors are of order $10^{-6}$. All correlator signs are preserved, resulting in identical raw and refined MaxCut solutions to those obtained classically.

\begin{table}[H]
\centering
\caption{Statevector transfer results at $\epsilon=0.02$ for the initialization with the highest raw empirical AR of each Gset instance. Reported metrics are post-selection success probability, state fidelity, TV distance, mean and maximum absolute correlator errors, sign agreement, and raw and refined empirical ARs. All correlator errors are relative to the classical Walsh representation, and the cut difference is zero in every instance.}
\label{tab:transfer_statevector}
\setlength{\tabcolsep}{3.5pt}
\resizebox{\linewidth}{!}{%
\begin{tabular}{lcccccccc}
\toprule
Instance & $P_{\rm succ}$ & fidelity & TV & corr. MAE & corr. max AE
& sign agreement & raw AR & refined AR \\
\midrule
G14 & $1.0\times10^{-4}$ & $0.999981$ & $1.4\times10^{-5}$ & $1.1\times10^{-6}$ & $5.5\times10^{-6}$ & $1.000$ & $0.9827$ & $0.9879$ \\
G23 & $1.0\times10^{-4}$ & $0.999949$ & $2.3\times10^{-5}$ & $1.3\times10^{-6}$ & $4.9\times10^{-6}$ & $1.000$ & $0.9898$ & $0.9957$ \\
G60 & $1.0\times10^{-4}$ & $0.999950$ & $2.5\times10^{-5}$ & $8.1\times10^{-7}$ & $3.5\times10^{-6}$ & $1.000$ & $0.9772$ & $0.9808$ \\
G64 & $1.0\times10^{-4}$ & $0.999953$ & $2.3\times10^{-5}$ & $7.0\times10^{-7}$ & $2.9\times10^{-6}$ & $1.000$ & $0.9330$ & $0.9491$ \\
\bottomrule
\end{tabular}%
}
\end{table}

These results confirm that the classically optimized Walsh coefficients can be transferred to quantum circuits without retraining, preserving the encoded correlators and MaxCut solutions in noiseless simulation. However, the post-selection probability at $\epsilon=0.02$ is only about $10^{-4}$, making this operating point unsuitable for practical finite-shot execution. The validation on real quantum hardware is therefore performed separately, using a smaller graph and a different operating point, as described in \cref{app:hardware_validation}.


\section{Conclusion}
\label{sec:conclusion}

We introduced a sparse Walsh-Pauli correlation encoding for MaxCut in which binary decision variables are represented by the signs of diagonal Pauli-string expectation values. By exploiting the equivalence between Pauli-$Z$ strings and Walsh operators, these correlators were expressed exactly as normalized XOR autocorrelations of a real coefficient vector. This formulation preserves the operational interpretation of PCE while allowing the optimization stage to be performed classically, without statevector construction or quantum expectation-value estimation.

The complete Walsh representation provided a reference of maximal diagonal expressivity and produced high-quality solutions with mean empirical ARs close to $0.98$. The subsequent support-scaling analysis showed that increasing the number of base masks did not yield a systematic improvement under the same optimizer and training budget. The single-base paired support, $\mathcal{A}=\{0\}\cup\mathcal{S}$ and $|\mathcal{A}|=m+1$, already recovered essentially the full-basis solution quality across all investigated sizes. Thus, progressively enlarging the support was not necessary to achieve comparable performance. Consistent with the sign-universality result in \cref{prop:sign_universality}, the single-base support can represent any binary assignment, while its ability to recover complete-basis solution quality under the tested optimization budget is established empirically.

These findings motivated the use of the single-base model on the public Gset benchmark, where it achieved mean refined ARs of $0.9828$, $0.9926$, $0.9781$, and $0.9369$ on G14, G23, G60, and G64, respectively, corresponding to a $5.1\times$-$9.4\times$ reduction in parameter count. The lower performance on the denser signed instance G64 indicates that this regime is more challenging under the configuration considered here and benefits more strongly from classical local refinement. However, since neither the hyperparameters nor the support size were specifically tuned for graphs with both positive and negative unit weights, this result should not be interpreted as an intrinsic limitation of the diagonal single-base model.

Beyond classical optimization, the trained coefficients were transferred to quantum circuits through commuting Walsh-diagonal rotations, ancilla-controlled interference, and post-selection. In statevector simulation, the transferred states reached fidelities above $0.99994$, reproduced the correlators with errors of order $10^{-6}$, and preserved all decoded signs. Separately, a compact instance with $15$ encoded variables was implemented on the \texttt{ibm\_fez} processor using six data qubits and one ancilla. Without readout mitigation or dynamical error-suppression techniques, the measured correlators achieved a sign agreement of $0.927\pm0.049$ relative to the corresponding noiseless finite-$\epsilon$ circuit, yielding raw and refined ARs of $0.917\pm0.068$ and $0.950\pm0.043$.

Despite this agreement, the hardware experiments also illustrate a practical limitation of the current transfer protocol. High-fidelity loading requires a small interference parameter, whereas the corresponding post-selection probability decreases quadratically. For the larger Gset instances, the combination of greater circuit depth, entangling-gate requirements, post-selection overhead, and hardware noise makes direct execution challenging. The quantum implementation should therefore be interpreted as a validation of parameter transfer and measurability on a compact instance, rather than as a demonstration of quantum computational advantage.

These limitations motivate future investigations of lower-depth controlled-Walsh synthesis, circuit distillation into hardware-efficient ansatzes, and transfer objectives that reproduce only the decision correlators rather than the complete function-state distribution. Further studies are also required to characterize the limitations of the single-base support on dense signed graphs and to determine whether larger paired supports, adaptive support constructions, or alternative diagonal representations can improve this regime without sacrificing the linear trainable-parameter scaling.

\section*{Acknowledgments}

This work was executed under the TIC26 -- Brazil Quantum Camp project,
funded within the scope of the Prioritized Informatics Programs and
Projects (PPI), Process No.\ 01245.008254/2025-22, under the
responsibility of the Ministry of Science, Technology and Innovation
(MCTI), with operational coordination by the Association for the
Promotion of Brazilian Software Excellence (SOFTEX), and executed by
CESAR and the Instituto de Pesquisas Eldorado.

\section*{Author Contributions}
C. A. do Amaral led the conceptualization, implementation, and analysis,
and wrote the manuscript. N. Legnazzi contributed to the implementation
and analysis. M. V. Reballo, M. Ritt, A. S. da Rosa Júnior, and
F. A. C. de Barros reviewed the manuscript and supervised the work.
\section*{Data Availability}
The Gset benchmark instances used in this study are publicly available~\citep{ye_gset}.
The random-graph bank and the hardware measurement data generated in this study are not
publicly available due to confidentiality restrictions of the
Instituto de Pesquisas Eldorado, but are available from the corresponding
author on reasonable request and subject to institutional approval.
\section*{Code Availability}
The source code and scripts developed for this study are proprietary to
the Instituto de Pesquisas Eldorado and are not publicly available.
Further details may be obtained from the corresponding author on reasonable
request, subject to institutional approval.
\section*{Competing Interests}
The authors declare no competing interests.

\include{appendix}
\bibliographystyle{unsrtnat}
\bibliography{references}

\end{document}

%% file: appendix.tex
\appendix
\section{Worked Example of Structured Walsh-Support Reduction}
\label{app:structured_support_example}

This appendix illustrates how paired support reduces the number of active coefficients without removing the correlators required to encode the variables. Consider a \(q=4\)-qubit register and three decision variables associated with the Hamming-weight-two masks
\begin{equation}
    S_1=(0011)_2=3,
    \qquad
    S_2=(0101)_2=5,
    \qquad
    S_3=(1001)_2=9.
    \label{eq:appendix_decision_masks}
\end{equation}

In the complete representation, the Walsh expansion contains \(2^q=16\) coefficients,
\begin{equation}
    f_{\mathbf a}(\mathbf x)
    =
    \sum_{\ell=0}^{15}
    a_\ell w_\ell(\mathbf x),
\end{equation}
and each correlator is evaluated as
\begin{equation}
    c_{S_i}(\mathbf a)
    =
    \frac{
        \displaystyle
        \sum_{\ell=0}^{15}
        a_\ell a_{\ell\oplus S_i}
    }{
        \displaystyle
        \sum_{\ell=0}^{15}
        a_\ell^2
    }.
    \label{eq:appendix_complete_correlator}
\end{equation}
For any nonzero mask \(S_i\), the \(16\) indices form eight unordered pairs \(\{\ell,\ell\oplus S_i\}\). The complete representation uses all of these pairs.

\subsection{Single-base support}

Consider first a single base mask,
\begin{equation}
    \mathcal B_1
    =
    \{(0000)_2\}.
\end{equation}
The construction in \cref{eq:structured_paired_support} gives
\begin{align}
    \mathcal A_1
    &=
    \mathcal B_1
    \cup
    \left\{
        b\oplus S_i:
        b\in\mathcal B_1,\;
        i=1,2,3
    \right\}
    \nonumber\\
    &=
    \{
        (0000)_2,
        (0011)_2,
        (0101)_2,
        (1001)_2
    \}
    \nonumber\\
    &=
    \{0,3,5,9\}.
    \label{eq:appendix_b1_support}
\end{align}
The number of active coefficients is therefore
\begin{equation}
    |\mathcal A_1|
    =
    m+1
    =
    4,
\end{equation}
instead of the \(16\) coefficients in the complete representation.

Defining
\begin{equation}
    D_1
    =
    a_0^2+a_3^2+a_5^2+a_9^2,
\end{equation}
the sparse correlators are
\begin{equation}
    c_{S_1}^{(\mathcal A_1)}
    =
    \frac{2a_0a_3}{D_1},
    \qquad
    c_{S_2}^{(\mathcal A_1)}
    =
    \frac{2a_0a_5}{D_1},
    \qquad
    c_{S_3}^{(\mathcal A_1)}
    =
    \frac{2a_0a_9}{D_1}.
    \label{eq:appendix_b1_correlators}
\end{equation}
The factor of two arises because the sum in \cref{eq:sparse_walsh_autocorrelation} contains both orientations of each pair,
\begin{equation}
    (0,S_i)
    \qquad\text{and}\qquad
    (S_i,0).
\end{equation}

In this example, the single-base support preserves exactly one of the eight possible pairs for each correlator. Each variable therefore remains explicitly parametrized, although the representation has fewer degrees of freedom than the complete model.

\subsection{Two-base support}

To increase expressivity, consider
\begin{equation}
    \mathcal B_2
    =
    \{
        (0000)_2,
        (1111)_2
    \}.
\end{equation}
The paired support becomes
\begin{align}
    \mathcal A_2
    &=
    \mathcal B_2
    \cup
    \left\{
        b\oplus S_i:
        b\in\mathcal B_2,\;
        i=1,2,3
    \right\}
    \nonumber\\
    &=
    \{
        (0000)_2,
        (0011)_2,
        (0101)_2,
        (0110)_2,
        (1001)_2,
        (1010)_2,
        (1100)_2,
        (1111)_2
    \}
    \nonumber\\
    &=
    \{0,3,5,6,9,10,12,15\}.
    \label{eq:appendix_b2_support}
\end{align}
In this case,
\begin{equation}
    |\mathcal A_2|
    =
    8
    =
    B(m+1),
\end{equation}
so half of the complete representation remains active.

With
\begin{equation}
    D_2
    =
    \sum_{\ell\in\mathcal A_2}a_\ell^2,
\end{equation}
the exact expression for the first correlator is
\begin{equation}
    c_{S_1}^{(\mathcal A_2)}
    =
    \frac{
        2\left(
            a_0a_3
            +
            a_5a_6
            +
            a_9a_{10}
            +
            a_{12}a_{15}
        \right)
    }{
        D_2
    }.
    \label{eq:appendix_b2_c1}
\end{equation}
Similarly,
\begin{equation}
    c_{S_2}^{(\mathcal A_2)}
    =
    \frac{
        2\left(
            a_0a_5
            +
            a_3a_6
            +
            a_9a_{12}
            +
            a_{10}a_{15}
        \right)
    }{
        D_2
    },
    \label{eq:appendix_b2_c2}
\end{equation}
and
\begin{equation}
    c_{S_3}^{(\mathcal A_2)}
    =
    \frac{
        2\left(
            a_0a_9
            +
            a_3a_{10}
            +
            a_5a_{12}
            +
            a_6a_{15}
        \right)
    }{
        D_2
    }.
    \label{eq:appendix_b2_c3}
\end{equation}

The \(B=2\) support preserves four XOR pairs for each variable, whereas the \(B=1\) support preserves only one. Increasing the number of base masks therefore introduces additional contributing pairs for adjusting each correlator.

\subsection{Interpretation}

In this example, the sequence of representations is
\begin{equation}
    \underbrace{16}_{\text{complete support}}
    \;\longrightarrow\;
    \underbrace{8}_{B=2}
    \;\longrightarrow\;
    \underbrace{4}_{B=1}
\end{equation}
active coefficients. The reduction does not replace the Walsh basis with another basis; rather, it restricts the expansion support while preserving index pairs separated by the decision masks.

The \(B=1\) construction guarantees that every correlator is parametrized, but it does not guarantee exact reproduction of an arbitrary complete-representation correlator. Increasing \(B\) adds XOR pairs and enlarges the model capacity. The central empirical question is therefore how small the support can be without a relevant loss in MaxCut solution quality.

\section{Complete-Basis Representability}
\label{sec:complete_basis_representability}
\label{app:walsh_representability}

\begin{proposition}[Complete representability of diagonal correlators]
\label{prop:complete_basis_representability}
Let \(\lvert\psi\rangle\) be a \(q\)-qubit state and let
\(\mathcal S=\{S_i\}_{i=1}^{m}\) be any set of diagonal PCE masks. There exists a complete real vector of Walsh coefficients \(\mathbf a\) such that
\begin{equation}
    \langle f(\mathbf a)\rvert
    \widehat W_{S_i}
    \lvert f(\mathbf a)\rangle
    =
    \langle\psi\rvert
    \widehat W_{S_i}
    \lvert\psi\rangle
    \label{eq:complete_representability_equality}
\end{equation}
for every \(i=1,\ldots,m\).
\end{proposition}

\begin{proof}
Define the probability distribution of \(\lvert\psi\rangle\) in the computational basis as
\begin{equation}
    p_\psi(\mathbf x)
    =
    \abs{
        \langle\mathbf x\vert\psi\rangle
    }^2
\end{equation}
and consider the real nonnegative function
\begin{equation}
    f_\psi(\mathbf x)
    =
    \sqrt{
        p_\psi(\mathbf x)
    }.
\end{equation}
Since \(\sum_{\mathbf x}p_\psi(\mathbf x)=1\), the state
\begin{equation}
    \lvert f_\psi\rangle
    =
    \sum_{\mathbf x}
    f_\psi(\mathbf x)
    \lvert\mathbf x\rangle
\end{equation}
is normalized. For any mask \(S\),
\begin{align}
    \langle f_\psi\rvert
    \widehat W_S
    \lvert f_\psi\rangle
    &=
    \sum_{\mathbf x}
    p_\psi(\mathbf x)
    w_S(\mathbf x)
    \nonumber\\
    &=
    \langle\psi\rvert
    \widehat W_S
    \lvert\psi\rangle.
    \label{eq:diagonal_statistics_equivalence}
\end{align}
Because Walsh functions form a complete basis, \(f_\psi\) has coefficients
\begin{equation}
    a_\ell
    =
    \frac{1}{2^q}
    \sum_{\mathbf x}
    f_\psi(\mathbf x)
    w_\ell(\mathbf x),
\end{equation}
which define a state \(\lvert f(\mathbf a)\rangle=\lvert f_\psi\rangle\).
\end{proof}

The proposition implies that, if a state \(\lvert\psi^\star\rangle\) produces an optimal vector of diagonal correlators
\begin{equation}
    \mathbf c^\star
    =
    \left(
        \langle\psi^\star\rvert
        \widehat W_{S_1}
        \lvert\psi^\star\rangle,
        \ldots,
        \langle\psi^\star\rvert
        \widehat W_{S_m}
        \lvert\psi^\star\rangle
    \right),
\end{equation}
then a complete coefficient vector \(\mathbf a^\star\) exists that reproduces \(\mathbf c^\star\) exactly. Since the relaxed spins, loss, and binary solution depend only on these correlators,
\begin{equation}
    c_{S_i}(\mathbf a^\star)
    =
    c_i^\star
    \quad\Longrightarrow\quad
    s_i(\mathbf a^\star)
    =
    s_i^\star,
    \qquad
    \mathcal L(\mathbf a^\star)
    =
    \mathcal L^\star.
\end{equation}

Thus, the complete Walsh representation has zero representability error for any physically realizable set of diagonal PCE correlators. Any suboptimality of the complete-basis optimizer therefore cannot be attributed to insufficient diagonal representability alone.

\section{Sign Universality of the Single-Base Support}
\label{app:sign_universality}

The single-base support guarantees that every correlator is parametrized. In fact it guarantees more: every binary assignment is representable.

\begin{proposition}[Sign universality of the single-base support]
\label{prop:sign_universality}
Let \(\mathcal S=\{S_i\}_{i=1}^{m}\) be the decision masks, all of Hamming weight \(k\), and let \(\mathcal A=\{0\}\cup\mathcal S\) be the single-base support. For every assignment \(\mathbf z\in\{-1,+1\}^{m}\) there is a coefficient vector \(\mathbf a\) supported on \(\mathcal A\) with
\begin{equation}
    \sign\left(c^{(\mathcal A)}_{S_i}(\mathbf a)\right)
    =
    z_i,
    \qquad
    i=1,\ldots,m.
    \label{eq:sign_universality}
\end{equation}
\end{proposition}

\begin{proof}
Let \(\delta>0\) and set \(a_0=1\) and \(a_{S_i}=\delta z_i\). In the numerator of \cref{eq:sparse_walsh_autocorrelation} for the mask \(S_i\), the indices \(\ell=0\) and \(\ell=S_i\) contribute \(a_0a_{S_i}+a_{S_i}a_0=2\delta z_i\). Any further contribution comes from an index \(\ell=S_j\) with \(j\neq i\) and \(S_j\oplus S_i\in\mathcal S\), and equals \(\delta^{2}z_jz_l\) for the mask \(S_l=S_j\oplus S_i\). Writing \(X_i\) for the sum of these terms, and using \(\sum_{\ell\in\mathcal A}a_\ell^{2}=1+m\delta^{2}>0\),
\begin{equation}
    c^{(\mathcal A)}_{S_i}(\mathbf a)
    =
    \frac{2\delta z_i+\delta^{2}X_i}{1+m\delta^{2}},
    \qquad
    \lvert X_i\rvert\leq m-1.
    \label{eq:sign_universality_correlator}
\end{equation}
The denominator is positive, and choosing $0<\delta<2/m$ ensures $|\delta^2 X_i|<2\delta$. Hence, the numerator has sign $z_i$, proving \cref{eq:sign_universality}.

If \(k\) is odd the condition on \(\delta\) can be dropped. The masks \(S_i\) and \(S_j\) both have weight \(k\), so \(S_j\oplus S_i\) has even weight and cannot equal a mask of odd weight \(k\). Hence \(X_i=0\) and \(c^{(\mathcal A)}_{S_i}(\mathbf a)=2\delta z_i/(1+m\delta^{2})\) for every \(\delta>0\).

The property carries over to every support \(\mathcal A'\supseteq\mathcal A\): setting the coefficients in \(\mathcal A'\setminus\mathcal A\) to zero leaves the denominator of \eqref{eq:sparse_walsh_autocorrelation} unchanged and adds only vanishing terms to its numerator, so all correlators, and with them the decoded assignment, are the same. In particular this holds for the supports of every \(B>1\), which contain the single-base support by construction.
\end{proof}

Two limitations should be noted. The proposition concerns signs only: it does not claim that the single-base support reproduces arbitrary correlator magnitudes, and indeed \cref{eq:sign_universality_correlator} shows that the constructed correlators are of order \(\delta\). It is also a statement about representability, not about reachability. The constructed coefficients place every relaxed spin \(\tanh(\alpha c_{S_i})\) close to zero, where the gradient of the quadratic MaxCut objective with respect to the relaxed spins also approaches zero. Therefore, representability does not guarantee that the optimizer will find such a solution. The empirical question addressed in this work therefore concerns the optimization dynamics and the geometry of the relaxed loss, not the ability of the single-base model to represent binary solutions.

\section{Gate-level implementation of the transfer circuit}
\label{sec:gate_implementation}

The optimized Walsh coefficients can be transferred directly to a quantum
circuit through the diagonal operator
\begin{equation}
\widehat{F}_{\mathbf{a}^{\star}}
=
\sum_{\ell\in\mathcal{A}}
a_{\ell}^{\star}\widehat{W}_{\ell},
\end{equation}
and the associated unitary
\begin{equation}
\widehat{U}_{\mathbf{a}^{\star},\epsilon}
=
\exp\left(
-i\epsilon
\widehat{F}_{\mathbf{a}^{\star}}
\right),
\label{eq:transfer_unitary}
\end{equation}
where \(\epsilon\) controls the finite-angle approximation used in the
state-preparation protocol. Since all Walsh operators are diagonal and
mutually commuting, the unitary can be factorized as~\citep{welch2014,zylberman2024}
\begin{equation}
\widehat{U}_{\mathbf{a}^{\star},\epsilon}
=
\prod_{\ell\in\mathcal{A}}
\exp\left(
-i\epsilon a_{\ell}^{\star}\widehat{W}_{\ell}
\right).
\label{eq:transfer_factorization}
\end{equation}
This factorization allows each nonzero Walsh coefficient to be implemented
independently using a parity-controlled \(Z\) rotation.

\begin{figure}[t]
    \centering
    \includegraphics[width=\columnwidth]{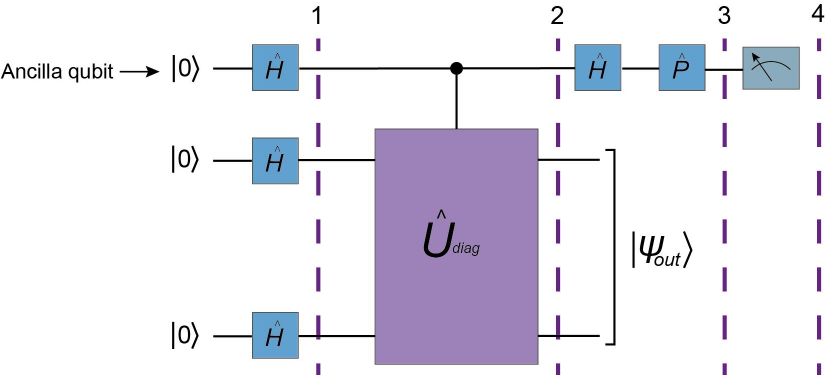}
    \caption{Author-generated schematic of the ancilla-interference and
        post-selection protocol used to transfer the optimized Walsh coefficients,
        based on the Walsh Series Loader construction of Zylberman and
        Debbasch~\citep{zylberman2024}. The protocol consists of four stages:
        preparation of a uniform superposition, application of the controlled
        diagonal unitary, interference on the ancilla, and ancilla measurement
        followed by post-selection.}
    \label{fig:transfer_circuit}
\end{figure}

The complete state-loading protocol is shown in
\cref{fig:transfer_circuit}. The notation in the figure follows the original
Walsh-loading construction: the operator denoted by
\(\widehat{U}_{\mathrm{diag}}\) corresponds here to
\(\widehat{U}_{\mathbf{a}^{\star},\epsilon}\), whereas the post-selected
state \(\ket{\psi_{\mathrm{out}}}\) corresponds to
\(\ket{\phi_{\epsilon}(\mathbf{a}^{\star})}\).

The circuit starts from the joint ancilla--data state
\begin{equation}
\ket{\psi_{0}}
=
\ket{0}_{a}\ket{0}^{\otimes q}.
\end{equation}
Hadamard gates are first applied to the ancilla and to all qubits of the
data register, producing
\begin{equation}
\ket{\psi_{1}}
=
\ket{+}_{a}\ket{+}^{\otimes q}
=
\frac{1}{\sqrt{2}}
\left(
\ket{0}_{a}
+
\ket{1}_{a}
\right)
\ket{+}^{\otimes q}.
\end{equation}
The diagonal transfer unitary is then applied conditionally on the ancilla,
yielding
\begin{equation}
\ket{\psi_{2}}
=
\frac{1}{\sqrt{2}}
\left[
\ket{0}_{a}\ket{+}^{\otimes q}
+
\ket{1}_{a}
\widehat{U}_{\mathbf{a}^{\star},\epsilon}
\ket{+}^{\otimes q}
\right].
\label{eq:controlled_transfer_state}
\end{equation}

Interference between the two branches is generated by applying a Hadamard
gate to the ancilla. The phase operation indicated by \(P\) in
\cref{fig:transfer_circuit},
\begin{equation}
P(\theta)
=
\operatorname{diag}(1,e^{i\theta}),
\end{equation}
is subsequently applied with \(\theta=-\pi/2\) to remove the relative factor \(i\) that appears
in the small-\(\epsilon\) expansion. The resulting state is
\begin{equation}
\begin{split}
\ket{\psi_{3}}
={}&
\frac{1}{2}
\ket{0}_{a}
\left(
I+
\widehat{U}_{\mathbf{a}^{\star},\epsilon}
\right)
\ket{+}^{\otimes q}
\\
&-
\frac{i}{2}
\ket{1}_{a}
\left(
I-
\widehat{U}_{\mathbf{a}^{\star},\epsilon}
\right)
\ket{+}^{\otimes q}.
\end{split}
\label{eq:interference_state}
\end{equation}
Therefore, measuring the ancilla and post-selecting the outcome \(a=1\)
leaves the data register in the normalized state
\begin{equation}
\ket{\phi_{\epsilon}(\mathbf{a}^{\star})}
=
\frac{
\left(
I-
\widehat{U}_{\mathbf{a}^{\star},\epsilon}
\right)
\ket{+}^{\otimes q}
}{
\left\|
\left(
I-
\widehat{U}_{\mathbf{a}^{\star},\epsilon}
\right)
\ket{+}^{\otimes q}
\right\|_{2}
},
\label{eq:finite_epsilon_state}
\end{equation}
up to an irrelevant global phase. The state is defined whenever \(P_{\mathrm{succ}}>0\), which fails only if the diagonal phases of \(\widehat{U}_{\mathbf{a}^{\star},\epsilon}\) are all multiples of \(2\pi\); by \cref{eq:success_probability_small_epsilon} this does not occur for the unit-normalized coefficients and the values of \(\epsilon\) used here.

To see how this state is related to the optimized Walsh function, expanding
the transfer unitary for small \(\epsilon\) gives
\begin{equation}
\widehat{U}_{\mathbf{a}^{\star},\epsilon}
=
I
-
i\epsilon
\widehat{F}_{\mathbf{a}^{\star}}
-
\frac{\epsilon^{2}}{2}
\widehat{F}_{\mathbf{a}^{\star}}^{2}
+
\mathcal{O}(\epsilon^{3}),
\end{equation}
and therefore
\begin{equation}
I-
\widehat{U}_{\mathbf{a}^{\star},\epsilon}
=
i\epsilon
\widehat{F}_{\mathbf{a}^{\star}}
+
\frac{\epsilon^{2}}{2}
\widehat{F}_{\mathbf{a}^{\star}}^{2}
+
\mathcal{O}(\epsilon^{3}).
\end{equation}
Consequently,
\begin{equation}
\ket{\phi_{\epsilon}(\mathbf{a}^{\star})}
\xrightarrow[\epsilon\rightarrow 0]{}
\frac{
\widehat{F}_{\mathbf{a}^{\star}}
\ket{+}^{\otimes q}
}{
\left\|
\widehat{F}_{\mathbf{a}^{\star}}
\ket{+}^{\otimes q}
\right\|_{2}
},
\label{eq:walsh_state_limit}
\end{equation}
showing that the post-selected circuit asymptotically prepares the quantum
state associated with the optimized Walsh function.

The probability of obtaining the required ancilla outcome is
\begin{equation}
P_{\mathrm{succ}}
=
\frac{1}{4}
\left\|
\left(
I-
\widehat{U}_{\mathbf{a}^{\star},\epsilon}
\right)
\ket{+}^{\otimes q}
\right\|_{2}^{2}.
\label{eq:success_probability}
\end{equation}
For small \(\epsilon\),
\begin{equation}
P_{\mathrm{succ}}
=
\frac{\epsilon^{2}}{4}
\left\|
\widehat{F}_{\mathbf{a}^{\star}}
\ket{+}^{\otimes q}
\right\|_{2}^{2}
+
\mathcal{O}(\epsilon^{4}),
\label{eq:success_probability_small_epsilon}
\end{equation}
which makes explicit the trade-off between the accuracy of the
finite-\(\epsilon\) approximation and the post-selection probability.

At the gate level, each nonzero Walsh mask
\(\ell\) defines the support
\begin{equation}
J_{\ell}
=
\left\{
r:\ell_{r}=1
\right\},
\qquad
h_{\ell}
=
\lvert J_{\ell}\rvert,
\end{equation}
such that
\begin{equation}
\widehat{W}_{\ell}
=
\bigotimes_{r\in J_{\ell}} Z_{r},
\end{equation}
acting as the identity on the qubits \(r\notin J_{\ell}\).
The corresponding factor in
\cref{eq:transfer_factorization} is
\begin{equation}
\widehat{U}_{\ell}
=
\exp\left(
-i\epsilon a_{\ell}^{\star}
\widehat{W}_{\ell}
\right).
\end{equation}

To implement this operation, one qubit \(t\in J_{\ell}\) is selected as
the parity target. A staircase of CNOT gates first computes the parity of
the remaining qubits in \(J_{\ell}\) onto \(t\). The single-qubit rotation
\begin{equation}
R_{z}\left(
2\epsilon a_{\ell}^{\star}
\right)
=
\exp\left(
-i\epsilon a_{\ell}^{\star}Z_{t}
\right)
\end{equation}
is then applied to the target, after which the CNOT staircase is reversed.
Before hardware-dependent routing, an uncontrolled Walsh factor of Hamming
weight \(h_{\ell}\) therefore requires
\begin{equation}
2(h_{\ell}-1)
\end{equation}
CNOT gates and one \(R_{z}\) rotation.

The state-loading protocol requires the controlled form of the complete
transfer unitary. For each Walsh factor, the parity computation and
uncomputation remain confined to the data register, whereas the central
\(R_{z}\) rotation is replaced by a controlled-\(R_{z}\). When the ancilla
is in \(\ket{0}\), the parity staircase and its inverse cancel. When the
ancilla is in \(\ket{1}\), the corresponding Walsh rotation is applied.
Using a standard decomposition of the controlled-\(R_{z}\) into two
CNOT gates and single-qubit rotations, the controlled factor requires
\begin{equation}
2(h_{\ell}-1)+2
=
2h_{\ell}
\end{equation}
two-qubit gates before coupling-map routing and native-gate decomposition.

\begin{figure}[t]
\centering
\includegraphics[width=0.75\linewidth]{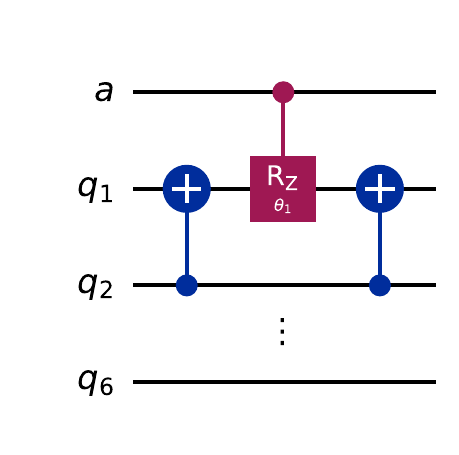}
\caption{Ancilla-controlled circuit realizing a single Walsh factor
\(\widehat{U}_{3}=\exp(-i\epsilon a_{3}^{\star}\,Z_{1}Z_{2})\) from the
factorization in \cref{eq:transfer_factorization}. Two CNOTs couple the masked
data qubits so that a single controlled-\(R_{z}\) applies the rotation conditioned on the
ancilla, after which the coupling is uncomputed. The rotation angle
\(\theta_{1}=2\epsilon a_{3}^{\star}\) is set by the classically trained Walsh
coefficient \(a_{3}^{\star}\), which is the only free parameter of the factor;
\(\epsilon\) is the fixed interference parameter shared by all factors. The
controlled-\(R_{z}\) is synthesized with two CNOTs, giving \(2h_{\ell}=4\)
two-qubit gates for this factor before routing.}
\label{fig:walsh_factor_circuit}
\end{figure}

As a concrete illustration, consider the decision mask
\(\ell=S_{1}=(000011)_{2}=3\) on the six-qubit register of the hardware
experiment, for which \(\widehat{W}_{3}=Z_{1}Z_{2}\). We emphasize that this
factor is one of the \(m\) independent terms of the product in
\cref{eq:transfer_factorization}, and not, by itself, the correlator of a
graph variable.
The factor to be realized is
\begin{equation}
\widehat{U}_{3}
=
\exp\!\left(-i\epsilon a_{3}^{\star}\,Z_{1}Z_{2}\right).
\end{equation}
Following the construction described above, this factor
requires two CNOT gates and one \(R_z\) rotation in its
uncontrolled form. The ancilla-controlled implementation
is illustrated in \cref{fig:walsh_factor_circuit}.

The constant Walsh component must be treated separately. Since
\begin{equation}
\widehat{W}_{0}=I,
\end{equation}
its uncontrolled contribution,
\begin{equation}
\exp\left(
-i\epsilon a_{0}^{\star}
\right)I,
\end{equation}
is only a global phase and would normally be discarded. In the controlled
construction, however, this phase becomes relative between the
\(\ket{0}_{a}\) and \(\ket{1}_{a}\) branches. It must therefore be retained and can be implemented directly as~\cite{pellini2025}
\begin{equation}
P\left(
-\epsilon a_{0}^{\star}
\right)
\end{equation}
on the ancilla, or equivalently by an \(R_{z}\) rotation up to a global
phase.

For the single-base support considered in this work,
\begin{equation}
\mathcal{A}
=
\{0\}
\cup
\mathcal{S},
\end{equation}
all \(m=\lvert\mathcal{S}\rvert\) nonconstant masks have Hamming weight \(k\).
The ideal number of two-qubit gates required by the controlled transfer
unitary is therefore at most
\begin{equation}
N_{\mathrm{2q}}
\mathrel{\leq}
2mk
=
\mathcal{O}(mk),
\label{eq:transfer_gate_scaling}
\end{equation}
rather than scaling with the complete Walsh-space dimension \(2^{q}\).
For the hardware instance considered here, \(m=15\) and \(k=2\), which
gives
\begin{equation}
N_{\mathrm{2q}}\mathrel{\leq}60
\end{equation}
two-qubit gates before hardware routing and basis synthesis. After
transpilation, the circuits contain \(119\) two-qubit gates, reflecting
the connectivity and native-gate constraints of the selected physical
qubits.

Finally, once the ancilla has been post-selected, all decision correlators
can be estimated from the same set of computational-basis measurements.
Because the observables are diagonal and mutually commuting, no additional
measurement basis or circuit is required for individual graph variables.
Given \(N_{\mathrm{ps}}\) post-selected bit strings \(x^{(t)}\), the
correlator associated with mask \(S_i\) is estimated as
\begin{equation}
\widehat{c}_{S_i}
=
\frac{1}{N_{\mathrm{ps}}}
\sum_{t=1}^{N_{\mathrm{ps}}}
(-1)^{S_i\cdot x^{(t)}}.
\label{eq:hardware_correlator_estimator}
\end{equation}
Thus, the same post-selected measurement data provide all correlators
required to decode the graph solution.

\section{Hardware Validation}
\label{app:hardware_validation}

\begin{figure}[H]
\centering
\includegraphics[width=0.7\linewidth]{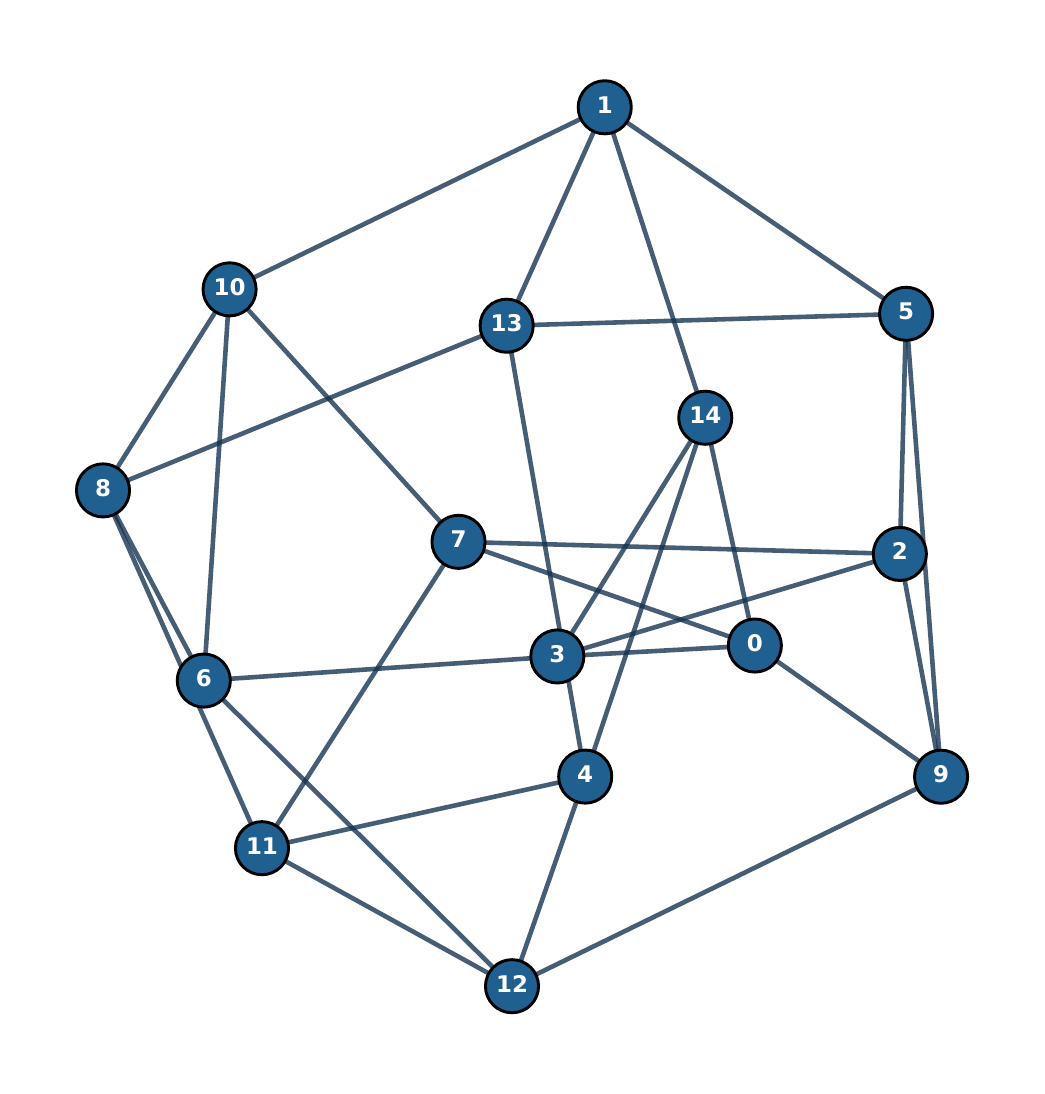}
\caption{Four-regular \(15\)-vertex instance used in the hardware experiment.}
\label{fig:hardware_graph}
\end{figure}

The experimental validation uses a compact instance compatible with the available device resources: a connected four-regular graph with \(15\) vertices and \(30\) edges, generated with seed \(20260715\), as shown in \cref{fig:hardware_graph}. Its exact optimum, \(C^\star=24\), was obtained by exhaustive enumeration, exploiting the global spin-flip symmetry of the MaxCut objective. The complete edge list is provided in \cref{tab:hardware_edge_list}. The \(15\) variables are associated with all weight-two masks on six qubits (\(\binom{6}{2}=15\)), and one additional qubit implements interference and post-selection.

The single-base model was trained from ten independent initializations, with seeds \(0,\ldots,9\), using Adam with learning rate \(\eta=10^{-3}\) for \(500\) epochs. Since the \(k=2\) hardware configuration lies outside the encoding orders considered in the random-graph calibration, the values \(\alpha=15\) and \(\beta=0.1\) were fixed manually for this hardware-validation experiment and were not selected by the calibration procedure of \cref{tab:hyperparameters_random_graphs}.

The resulting ten trained coefficient vectors were frozen and transferred to the circuit. For hardware execution, the interference parameter was fixed at \(\epsilon=0.6\) to obtain a substantially larger post-selection yield. This value was not selected by optimization of the MaxCut objective and is not interpreted as an optimized transfer parameter. Because the larger \(\epsilon\) introduces a finite-angle distortion relative to the ideal Walsh state, all hardware-level comparisons are made against the corresponding exact noiseless finite-\(\epsilon\) circuit.

The ten trained coefficient vectors were executed on \texttt{ibm\_fez} using Qiskit~\citep{qiskit2019}, with \(32{,}768\) shots requested per circuit. The circuits were transpiled using optimization level \(2\) with SABRE routing. No error mitigation or dynamical decoupling was applied.

The correlators estimated from post-selected measurements achieve a sign agreement of \(0.927\) relative to the exact noiseless finite-\(\epsilon\) statevector, and the decoded solution reaches raw and refined empirical ratios of \(0.917\) and \(0.950\), respectively (\cref{tab:hardware}). The hardware-derived raw cut matched the corresponding noiseless finite-\(\epsilon\) cut in 7 of the 10 runs, while the refined cut matched its noiseless counterpart in all 10 runs. These results show that the transferred correlators can be measured on quantum hardware without error mitigation, while also quantifying the remaining noise-induced degradation.

State fidelity is not reported for the hardware experiment because the experiment provides samples from the post-selected computational-basis distribution rather than a full reconstruction of the prepared quantum state. Estimating the state fidelity would require quantum-state tomography or an additional fidelity-estimation protocol. We therefore assess the hardware transfer using quantities directly accessible from measurement: the post-selection probability, TV distance, decision-correlator errors, and correlator-sign agreement. The dispersions reported in \cref{tab:hardware} are across the ten trained coefficient vectors. With approximately \(5{,}500\) post-selected shots, the standard error of an individual Pauli correlator estimate is bounded by \(0.014\).

\begin{table}[H]
\centering
\caption{Hardware validation of the \(m=15\) instance on \texttt{ibm\_fez} (\(\epsilon=0.6\), \(32{,}768\) shots, of which about \(5{,}500\) survive post-selection, \(10\) initializations). All distribution- and correlator-level errors are evaluated relative to the exact noiseless finite-\(\epsilon\) circuit with the same trained coefficients, not to the ideal Walsh target; empirical ARs are reported as mean \(\pm\) standard deviation.}
\label{tab:hardware}
\begin{tabular}{lc}
\toprule
Metric & Value \\
\midrule
Measured post-selection rate & $0.169 \pm 0.004$ \\
TV distance to noiseless       & $0.253 \pm 0.011$ \\
Correlator MAE to noiseless                 & $0.141 \pm 0.010$ \\
Sign agreement with noiseless                 & $0.927 \pm 0.049$ \\
Raw AR                                 & $0.917 \pm 0.068$ \\
Refined AR                             & $0.950 \pm 0.043$ \\
Noiseless-cut match (raw)                 & $7/10$ \\
Noiseless-cut match (refined)             & $10/10$ \\
\bottomrule
\end{tabular}
\end{table}
The measured post-selection rate ($0.169$) is nearly twice the noiseless prediction ($0.086$), indicating that hardware noise also affects the ancilla post-selection statistics. The increased acceptance rate should therefore not be interpreted as improved state-preparation efficiency.

\begin{figure}[H]
\centering
\includegraphics[width=0.75\linewidth]{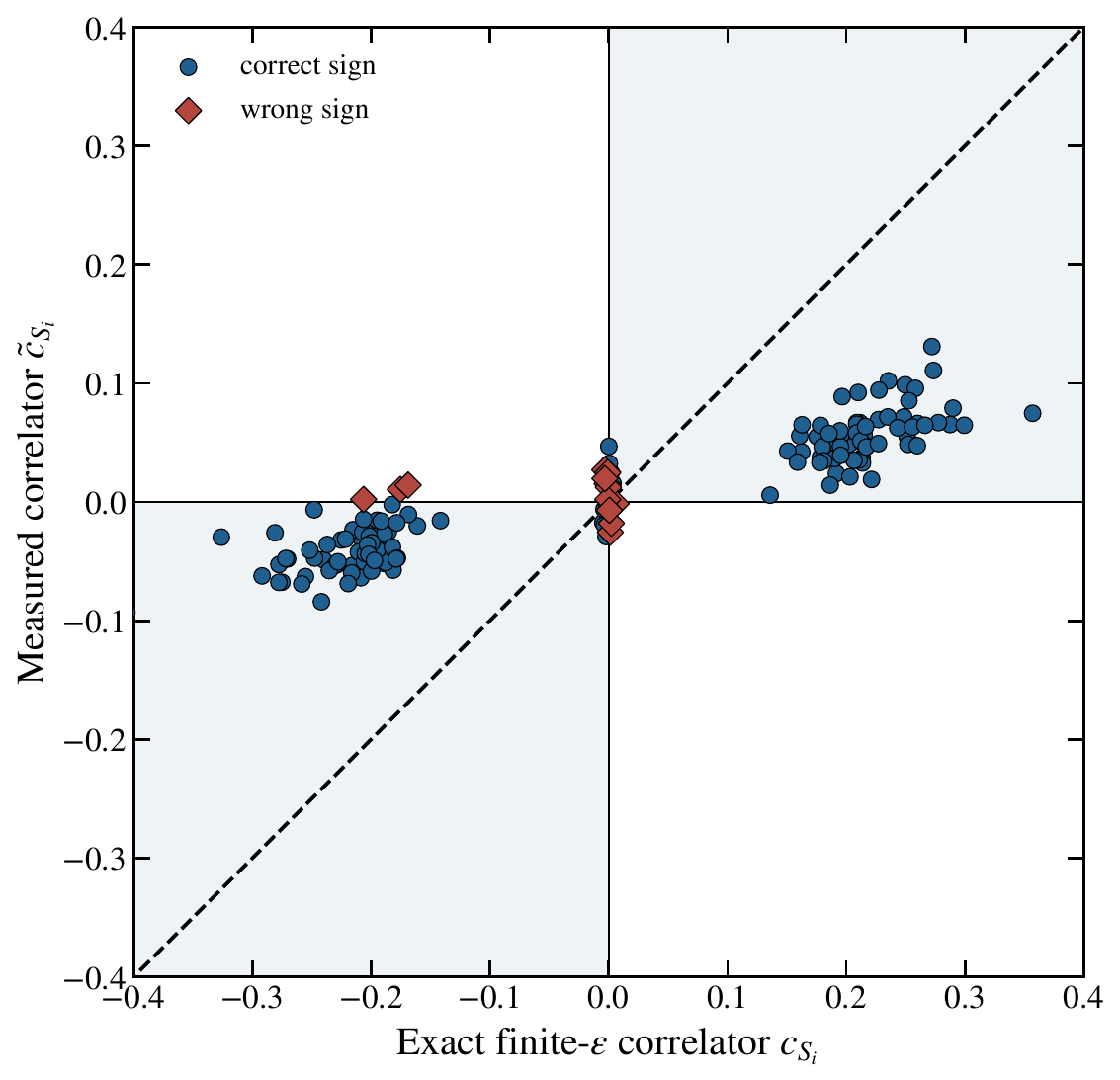}
\caption{Decision-mask correlators measured on hardware versus their exact values for the \(15\) masks and \(10\) initializations of the \(m=15\) instance. The dashed line denotes exact equality and shaded quadrants correspond to the correct sign.}
\label{fig:hardware_correlators}
\end{figure}

\cref{fig:hardware_correlators} compares the hardware correlators mask by mask with their exact values across the \(15\) masks and \(10\) initializations. Although the hardware distribution is substantially distorted (\(D_{\mathrm{TV}}=0.253\pm0.011\)), this distribution-level error does not translate into a comparable loss of task-relevant information: the Walsh correlator magnitudes are generally attenuated, while their signs remain correct in \(92.7\%\) of the cases. This distinction matters because the MaxCut decoding depends on the correlator signs rather than on reproducing the full probability distribution. The robustness is not a generic consequence of a moderate TV distance: the bound \(\lvert c_S(p)-c_S(q)\rvert\leq 2D_{\mathrm{TV}}=0.506\) of \cref{eq:corr_tv_bound} would, in the worst case, admit sign inversions for correlators with \(\lvert c_S\rvert\lesssim 0.5\). The observed errors are predominantly attenuating, with measured correlators generally smaller in magnitude than their exact counterparts, and the sign disagreements are concentrated where the exact correlator is already close to zero.

\begin{figure}[!h]
\centering
\includegraphics[width=0.85\linewidth]{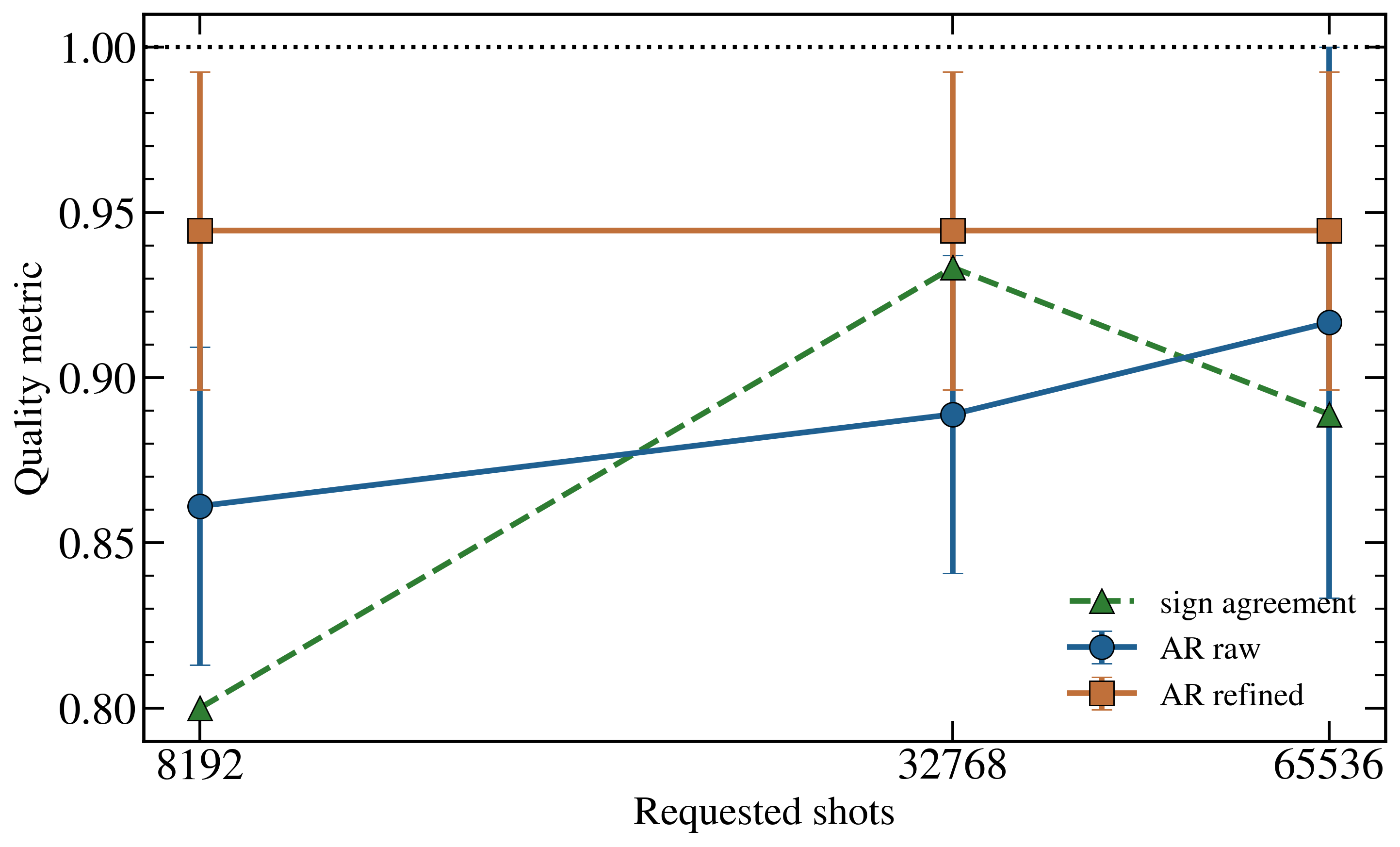}
\caption{Hardware validation of the \(m=15\) instance: raw and refined empirical ARs and correlator-sign agreement relative to the exact noiseless finite-\(\epsilon\) statevector, as functions of the number of requested shots. The dotted line denotes the exact optimum (\(\mathrm{AR}=1\)).}
\label{fig:hardware_shots}
\end{figure}

To assess the effect of the sampling budget on the hardware results, we performed separate hardware submissions with \(8{,}192\), \(32{,}768\), and \(65{,}536\) requested shots, using the same three trained coefficient vectors for each shot count, as shown in \cref{fig:hardware_shots}.

Across the three coefficient vectors used in the independent hardware submissions, the raw AR is higher at the larger requested-shot budgets. The correlator-sign agreement improves from \(8{,}192\) to \(32{,}768\) shots but is not monotonic at the largest budget. Because each point is based on only three trained coefficient vectors and separate hardware jobs, this sweep is interpreted descriptively rather than as a statistically established shot-scaling law. The refined AR remains comparatively stable because the final local-search step compensates for part of the residual variation in the measured correlators.

The resources of the transpiled circuits are summarized in \cref{tab:hardware_resources}. Each circuit uses seven physical qubits, has an ISA depth between \(264\) and \(268\), and contains \(119\) two-qubit gates.

\begin{table}[H]
\centering
\caption{Resources of the circuits transpiled for the \(m=15\) instance on \texttt{ibm\_fez}, over the \(10\) initializations. Quantities that vary across circuits are reported as mean [min--max].}
\label{tab:hardware_resources}
\begin{tabular}{lc}
\toprule
Resource & Value \\
\midrule
Active physical qubits       & $7$ \\
ISA depth                    & $265.6\,[264\text{--}268]$ \\
Two-qubit depth              & $98$ \\
Two-qubit gates              & $119$ \\
Circuit size                 & $668.4\,[666\text{--}673]$ \\
Circuit duration             & $10.86\,\mu\mathrm{s}$ \\
Maximum two-qubit error      & $2.39\times10^{-3}$ \\
Maximum readout error        & $1.51\times10^{-2}$ \\
\bottomrule
\end{tabular}
\end{table}

The reported two-qubit and readout error rates are based on device calibration data associated with the experiment and should not be interpreted as direct measurements of the errors incurred during circuit execution.

Taken together, the hardware results demonstrate that the classically trained sparse Walsh-Pauli representation can be transferred to a quantum processor and its decision correlators recovered directly from measurement. Despite the finite circuit depth and hardware noise, the measured correlators preserve the correct signs in most cases, yielding a mean raw empirical AR of \(0.917\). After applying the same local-search procedure used for the Gset and transfer results, the mean AR reaches \(0.950\), and the refined hardware cuts match their noiseless finite-\(\epsilon\) counterparts in all ten runs.

\section{Hardware-Instance Specification}
\label{app:hardware_instance}

The graph used for the real-device validation is the connected four-regular
\(15\)-vertex instance introduced in \cref{app:hardware_validation}. It was
generated with seed \(20260715\), contains \(30\) unit-weight edges, and has
exact optimum \(C^\star=24\), obtained by exhaustive enumeration.

\begin{table}[H]
\centering
\caption{Edge list of the four-regular \(15\)-vertex graph used in the hardware experiment. All edge weights are equal to one.}
\label{tab:hardware_edge_list}
\begin{tabular}{cccccc}
\toprule
\multicolumn{6}{c}{Edges} \\
\midrule
$(0,3)$  & $(0,7)$  & $(0,9)$  & $(0,14)$ & $(1,5)$  & $(1,10)$ \\
$(1,13)$ & $(1,14)$ & $(2,3)$  & $(2,5)$  & $(2,7)$  & $(2,9)$  \\
$(3,6)$  & $(3,14)$ & $(4,11)$ & $(4,12)$ & $(4,13)$ & $(4,14)$ \\
$(5,9)$  & $(5,13)$ & $(6,8)$  & $(6,10)$ & $(6,12)$ & $(7,10)$ \\
$(7,11)$ & $(8,10)$ & $(8,11)$ & $(8,13)$ & $(9,12)$ & $(11,12)$ \\
\bottomrule
\end{tabular}
\end{table}